\documentclass[12pt,prd,onecolumn,superscriptaddress,altaffilletter,amssymb,showpacs,nofootinbib]{revtex4}
\usepackage[dvips]{graphicx}
\usepackage{amsmath}
\usepackage{graphicx}
\usepackage{amsmath,amssymb}        
\usepackage[draft=false]{hyperref}
\usepackage{threeparttable}
\newcommand{\be}{\begin{equation}}
\newcommand{\ee}{\end{equation}}
\newcommand{\bea}{\begin{eqnarray}}
\newcommand{\eea}{\end{eqnarray}}
\newcommand{\vphi}{\varphi}
\newtheorem{theorem}{Theorem}
\newtheorem{thm}{Theorem}[section]

\newtheorem{lem}[thm]{Lema}

\newtheorem{prop}{Proposition}
\newtheorem{defn}{Definition}

\def\CQG{{Class. Quant. Grav. } }

\def\PL{{Phys. Lett.} }

\def\PRTS{{Physics Reports} }

\begin{document}
\title[On the past asymptotic dynamics of coupled dark energy]{On the Past Asymptotic Dynamics of Non-minimally Coupled Dark Energy}
\author{Genly Leon}
\address{Department of Mathematics, Universidad Central de Las Villas, Santa Clara CP 54830, Cuba}
 \email{genly@uclv.edu.cu}

\date{\today}

\begin{abstract}

We apply dynamical systems techniques to investigate cosmological
models inspired in scalar-tensor theories written in the Einstein
frame. We prove that if the potential and the coupling function
are sufficiently smooth functions, the scalar field almost always
diverges into the past. The dynamics of two important invariant
sets is investigated in some detail. By assuming some regularity
conditions for the potential and for the coupling function, it is
constructed a dynamical system well suited to investigate the
dynamics where the scalar field diverges, i.e. near the initial
singularity. The critical points therein are investigated and the
cosmological solutions associated to them are characterized. We
find that our system admits scaling solutions. Some examples are
taken from the bibliography to illustrate the major results. Also
we present asymptotic expansions for the cosmological solutions
near the initial space-time singularity, which extend in a way
previous results of other researchers.

\end{abstract}

\maketitle

\section{Introduction}

It is believed that the field responsible for early time inflation
as well as the field driving the current accelerated expansion is
in the form of an scalar field (see \cite{sncmbsdss} for the
observational status of the acceleration of expansion). Theories
including scalar fields, such as scalar-tensor theories (STT) of
gravity \cite{Brans:1961sx,STT}, can be supported by fundamental
physical theories like superstring theory \cite{Green:1996bh}.
Otherwise, scalar fields can be viewed merely as convenient
heuristic models to elucidates certain (qualitative) dynamical
features of the early and/or the late time universe \footnote{See
\cite{Gonzalez:2004dh} and references therein  for the analysis of
a scalar field responsible for both the early and the late time
inflationary expansion in the context of brane cosmology.}.
Quintessential dark energy (DE) models \cite{quintessence}, for
instance, are (heuristical proposals) described by an ordinary
scalar field minimally coupled to gravity. A wise selection of the
scalar field self-interacting potentials can drive the current
accelerated expansion. Other scalar field models have been treated
in the literature (see for instance, the reviews
\cite{SahniCopeland:2006wr}).

The natural generalizations to models of quintessence evolving
independently from the background matter are models which exhibit
non-minimal coupling between both components. Experimental tests
in the Solar system impose severe restrictions on the possibility
of non-minimal coupling between DE and ordinary matter fluids
\cite{BarrowandParsons}. However, we argue that, due to the
unknown nature of dark matter (DM), it is possible to have
additional (non gravitational) interactions between the DE and the
DM components. This argument does not enter in conflict with the
experimental data, but, when the stability of dark energy
potentials in quintessence models is considered, the dark
matter-dark energy coupling might be troublesome \cite{doran}.

The effective interaction dark energy-dark matter appears when we
apply conformal transformations in scalar-tensor theories (STT)
\footnote{See the reference \cite{Faraoni:1998qx} for applications
of conformal transformations in both relativity and cosmology.}.
About these theories, it is known that they survive several
observational tests including Solar System tests
\cite{solarsystem} and Big-Bang nucleosynthesis constraints
\cite{nucleosinthesis}. Its simplest proposal is the Brans-Dicke
theory (BDT) \cite{Brans:1961sx}, in which a scalar field, $\chi,$
acts as the source for the gravitational coupling with a varying
Newtonian 'constant'  $G\sim \chi^{-1}.$ More general STT with a
non-constant BD parameter $\omega(\chi),$ and non-zero
self-interaction potential $V(\chi),$ have been formulated, and
also survive astrophysical tests \cite{BarrowandParsons}.

The action for a general class of STT, written in the so-called
Einstein frame (EF), is given by \cite{Kaloper1997sh}:

\begin{align}&S_{EF}=\int_{M_4} d{ }^4 x \sqrt{|g|}\left\{\frac{1}{2} R-\frac{1}{2}(\nabla\phi)^2-V(\phi)+\chi(\phi)^{-2}
\mathcal{L}_{matter}\right\}\label{eq1}
\end{align} In this equation $R$ is the curvature scalar, $\phi$ is the a scalar field, related via conformal
transformations with the dilaton field, $\chi.$ $V(\phi)$ is the quintessence self-interaction potential,
$\chi(\phi)^{-2}$ is the coupling function,  $\mathcal{L}_{matter}(\mu,\nabla\mu,\chi(\phi)^{-1}g_{\alpha \beta})$ is the
 matter Lagrangian, $\mu$ is a collective name for the matter degrees of freedom.

By considering the conformal transformation $\bar{g}_{\alpha \beta}=\chi(\phi)^{-1}g_{\alpha \beta}$ and defining the
Brans-Dicke coupling 'constant' $\omega(\chi)$ in such way that $d\phi=\pm \sqrt{\omega(\chi)+3/2}\chi^{-1} d\chi$ and
recalling $\bar{V}(\chi)=\chi^2 V(\phi(\chi))$ the action (\ref{eq1}) can be written in the Jordan frame (JF) as
(see \cite{Coley:2003mj}):

\begin{align}& S_{JF}=\int_{M_4} d{ }^4 x \sqrt{|\bar{g}|}\left\{\frac{1}{2}\chi \bar{R}-\frac{1}{2}\frac{\omega(\chi)}
{\chi}(\bar{\nabla}\chi)^2-\bar{V}(\chi)
+\mathcal{L}_{matter}(\mu,\nabla\mu,\bar{g}_{\alpha \beta})\right\}\label{eq1JF}
\end{align}

Both frames are both formally and physically equivalent
\cite{equivalent}. This fact removes previous doubts and fully
establishes equivalence at the classical level; however, this does
not guarantee physical equivalence at the quantum level
\cite{quantum}.

By making use of the conformal equivalence between the Einstein
and Jordan frames we can find, for example, that the theory
formulated in the EF with the coupling function $\chi(\phi)=\chi_0
\exp((\phi-\phi_0)/\varpi),\; \varpi\equiv \pm\sqrt{\omega_0+3/2}$
and potential $V(\phi)=\beta
\exp({(\alpha-2){\varpi}/(\phi-\phi_0)})$ corresponds to the
Brans-Dicke theory (BDT) with a power law potential, i.e.,
$\omega(\chi)=\omega_0,\; \bar{V}(\chi)=\beta \chi^\alpha.$ Exact
solutions with exponential couplings and exponential potentials
(in the EF) were investigated in \cite{Gonzalez2006cj}.

In the STT given by (\ref{eq1JF}), the energy-momentum of the matter fields is separately conserved. However, when written
 in the EF (\ref{eq1}), this is no longer the case, although the overall energy density is conserved. In fact in the EF
 we find that

$$Q_\beta\equiv\nabla^\alpha T_{\alpha \beta}=-\frac{1}{2}T\frac{\partial_\phi\chi(\phi)}{\chi(\phi)}\nabla_{\beta}\phi,\;
 T=T^\alpha_\alpha$$  where $$T_{\alpha \beta}=-\frac{2}{\sqrt{|g|}}\frac{\delta}{\delta g^{\alpha \beta}}\left\{\sqrt{|g|}
 \chi^{-2}\mathcal{L}(\mu,\nabla\mu,\chi^{-1}g_{\alpha \beta})\right\}.$$

In the present investigation we study FRW space-times with flat
spatial slices, i.e., we consider the line element:
\begin{align}d s^2 = -d t^2 +a(t)^2\left({d r^2}+r^2\left(d \theta^2+\sin^2\theta d \varphi^2\right)\right).
\label{flatFRW}
\end{align} We use a system of units in which $8\pi G=c=\hbar=1.$

The simplest way of modelling matter and energy in the Universe is
to consider that the energy-momentum tensor $T_{\alpha\,\beta}$ is
in the form of a perfect fluid
$$T^\alpha_\beta=\text{diag} \left(-\rho,p,p,p\right),$$ where $\rho$ and
$p$ are respectively the isotropic energy density and the
isotropic pressure (consistently with FRW metric, pressure is
necessarily isotropic \cite{trodden}). For simplicity we will
assume a barotropic equation of state $p=(\gamma-1)\rho.$

In this case, the vector field $Q_\mu=$ has only one non-null
component, the time component, so it can be written as
$Q_\beta=(Q_0,0,0,0).$  $Q_0$ can be considered in some way as a
rate of energy exchange between the scalar field and the
background.

One of the first papers to take seriously the possibility of
interaction in scalar field cosmologies, from the dynamical system
perspective, was \cite{Billyard:2000bh}. In that paper we can find
a review on the subject. There it was investigated the interaction
terms (in the flat FRW geometry) $\delta=-\alpha\dot\phi\rho$ and
$\delta=\alpha\rho H,$ where $\alpha$ is a constant, $\phi$ is the
scalar field, $\rho$ is the energy density of background matter
and $H$ stands for the Hubble parameter. The first choice
corresponds to an exponential coupling function $\chi(\phi)=\chi_0
\exp\left(2 \alpha \phi/(4-3\gamma)\right).$ The second case
corresponds to the choice $\chi=\chi_0 a^{-2\alpha/(4-3\gamma)}$
(and then, $\rho\propto a^{\alpha-3\gamma}$), where $a$ denotes
the scale factor of the Universe (recall that the former
derivations are only valid in a flat FRW model). Other
phenomenological coupling functions were studied elsewhere. We
want to draw the attention of the reader to a physically well
motivated approach to the coupling function in
\cite{Boehmer:2008av}. In that paper it was investigated a
coupling term of the form $\delta=-\alpha \rho,$ where $\alpha$ is
a constant ($\Gamma$ in their notation). As commented in that
reference, if $\alpha>0$, the model can describe either the decay
of dark matter into radiation, the decay of the curvaton field
into radiation or the decay of dark matter into dark energy (see
section III of \cite{Boehmer:2008av} for more information and for
useful references). In the reference \cite{viableSTT}, the authors
construct a family of viable scalar-tensor models of dark energy
(which includes pure $f(R)$ theories and quintessence). They
consider a  coupling between the scalar field and the
non-relativistic matter in the Einstein frame of the type -in our
notation- $\chi(\phi)=e^{-2 Q \phi}$, with $Q$ constant. By
investigating a phase space the authors obtain that the model
posses a phase of late-time acceleration preceded by a standard
matter era, while at the same time satisfying the local gravity
constraints (LGC).  In fact, by studying the evolution of matter
density perturbations and employing them, the authors place bounds
on the coupling of the order $|Q|<2.5 \times 10^{-3}$ (for the
massless case). By a chameleon mechanism the authors show that
these models can be made compatible with LGC
even when $|Q|$ is of the order of unity if the scalar-field
potential is chosen to have a sufficiently large mass in the
high-curvature regions.

About the dynamics of coupled dark energy, it was found (see
\cite{Coley:2003mj} and references therein) that typically at
early times ($t\rightarrow 0$) the BDT solutions are approximated
by the vacuum solutions and at late times ($t\rightarrow\infty$)
by matter dominated solutions, in which the matter is dominated by
the BD scalar field (denoted by $\chi$ in the Jordan frame). Exact
perfect fluid solutions in STT of gravity with a non-constant BD
parameter $\omega(\chi)$ have been obtained by various authors
(see \cite{coley2}). Coupled quintessence was investigated also in
\cite{dynamical} by using dynamical systems techniques.

In order to classify the global behavior of the solutions of
(\ref{eq1}) it is required a detailed knowledge of the form of the
scalar field potential (and of the coupling function $\chi$).
However, up to the present, there exist no consensus about the
specific functional form of $V(\phi)$ (and of $\chi(\phi)$). As a
consequence it would be of interest to classify the dynamical
behavior of solutions without the prior of specifying the
functional form of the potential function (and of the coupling
function). In the literature on General Relativity (GR) several
attempts have been made in this more general direction. The
dynamical systems techniques have proven very useful to do so
\cite{Foster1998sk,Miritzis2003ym,arbitrary}. In this respect we
know the dynamical behavior of scalar field space-times for a wide
class of non-negative potentials (within Einstein gravity (EG))
\footnote{In the reference \cite{Hertog}, the nonnegativity of the
potential is relaxed and new results within the context of EG
generalizing those in \cite{Miritzis2003ym} are obtained.}.

In this investigation we want to study, from the dynamical systems
point of view, a phenomenological model inspired in a STT with
action (\ref{eq1}), where  the matter and the (quintessence)
scalar field are coupled in the action (\ref{eq1}) through the
scalar tensor metric  $\chi(\phi)^{-1}g_{\alpha \beta}$
\cite{Kaloper1997sh}.  \footnote{In \ref{appendixA} it is provided a summary of the main results of the theory of dynamical systems that are used in this paper.} We consider arbitrary functional forms for
the self-interaction potential and the coupling function for the
scalar field $\phi.$ When we take the conformal transformation
allowing writing the action in the JF as in (\ref{eq1JF}) the
coupling function $\chi$ should be interpreted as the dilation
(BD) field and the corresponding $\omega(\chi)$ as the varying BD
parameter.

By employing Hubble-normalized dynamical variables in addition to
the scalar field we find, by investigating the phase space, that
the scalar field almost always diverges into the past, allowing us
to identify this regime with the physical region on a vicinity of
the initial space-time singularity. For finite values of the
scalar field we find that the late time (early time) 
attractor is associated with the minimum of the logarithm of
the potential (coupling) function.

By assuming some general regularity conditions for the potential
and for the coupling function when $\phi\rightarrow\infty,$ and
using the formalism developed in \cite{Foster1998sk}, we are able
to explore the phase space corresponding to this limit. We are
able to compute the critical points corresponding to that limit
and to characterize the cosmological solutions associated to them.
Scaling solutions do arise in this regime. Also we are able to
characterize the initial singularity in STT. In \ref{appendixA} an example taken from literature is given to illustrate 
the formalism developed for analyzing the region where the scalar fields diverge.

\section{The model}

In the model described by the action (\ref{eq1}), the background
energy density does not necessarily corresponds to a Dark Matter
component and, analogously, the scalar field energy density does
not necessarily corresponds to a Dark Energy component. However,
as done many times in the literature (see for instance, \cite{Gonzalez2006cj,quiros}), we assume that $\phi$ is a
quintessence scalar field, which is coupled metrically to a
background of a perfect fluid. This model is viable
phenomenologically. The possibility of a universal coupling of
dark energy to all sorts of matter, including baryons (but
excluding radiation) is studied in \cite{Chimento2003iea}.

\subsection{The field equations}

By using the line element (\ref{flatFRW}), the Einstein's field equations (derived by varying the action (\ref{eq1})) are:  a) the Raychaudhuri equation

\begin{equation}
\dot H=-\frac{1}{2}\left(\gamma\, \rho +\dot\phi^2\right),\label{Raych}
\end{equation}
where the dot denotes derivative with respect the cosmic time $t$, $\rho$ is the energy density of dark matter, b) the
Friedmann equation

\begin{equation}
3 H^2= \frac{1}{2}\dot\phi^2+V(\phi)+\rho, \label{Fried}
\end{equation}
and the continuity equation

\begin{equation}
\dot\rho=-3\,\gamma\, H\, \rho-\frac{1}{2}\,\left(4-3\gamma\right)\,\rho\,\dot\phi\,\frac{\chi'(\phi)}{\chi(\phi)}
\label{Cont}.
\end{equation}

This equation can be integrated in quadratures to give the useful
relation $$\rho=\rho_0 a^{-3\gamma}\chi^{-2+3\gamma/2}.$$

 The equation of motion of the scalar field (written as two
differential equations, one for $\phi$ and the other for $\dot
\phi$) reads

\begin{equation}
\frac{d \phi}{d t}=\dot \phi \label{motion1}
\end{equation}

\begin{equation}
\frac{d \dot\phi}{d t}=-3\,H\, \dot\phi-V'(\phi)+\frac{1}{2}\left(4-3\gamma\right)\,\rho\,\frac{\chi'(\phi)}{\chi(\phi)}
\label{motion2}
\end{equation} where the coma denotes the derivative with respect to the scalar field.

Observe that, if $\chi=const.,$ the equations for the minimally
coupled theory are recovered.

From the equations (\ref{Raych}, \ref{Cont}, \ref{motion1},
\ref{motion2}) we see that $(H, \rho, \phi,
\dot\phi)\in\mathbb{R}^4$ remains in the hypersurface defined by
the restriction (\ref{Fried}). Thus, defining an autonomous system
in the phase space

\begin{equation}
\Omega=\left\{(H, \rho, \phi, \dot\phi)\in\mathbb{R}^4: 3H^2=\frac{1}{2}\dot\phi^2+V(\phi)+\rho \right\} \label{eq9}
\end{equation}

We shall consider the following general assumptions: $V(\phi)\in C^3$ and $V(\phi)\geq 0,$  $\chi(\phi)\in
C^3$ with $\chi(\phi)> 0,$  $\rho\geq 0,$ $0<\gamma<2,$
$\gamma\neq 4/3$ instead of considering specific choices for both the potential and the coupling function.

\section{Qualitative analysis on the Hubble normalized state space}\label{Qualitative}

In this section we rewrite equations (\ref{Raych}, \ref{Cont},
\ref{motion1}, \ref{motion2}) as an autonomous system defined on a
state space by introducing Hubble-normalized variables. This
variables satisfy an inequality arising from the Friedmann
equation (\ref{Fried}). We analyse the cosmological model by
investigating the flow of the autonomous system in a phase space
by using dynamical systems tools. In order to make the paper
self-contained we offer in \ref{appendixA} some terminology and
results from the theory of dynamical systems that we use in the
demonstration of our main results.

\subsection{Normalized variables}

In order to analyze the initial singularity and the late time
behavior it is convenient to normalize the variables, since in the
vicinity of a hypothetical initial singularity physical variables
would typically diverge, whereas at late times they commonly
vanish \cite{Wainwright2004cd}.

Let us introduce the following normalized variables

\be x=\frac{1}{H},\,y=\frac{\dot\phi}{\sqrt{6}H},\,z=\frac{\sqrt{\rho}}{\sqrt{3} H}\label{vars}\ee and the time
coordinate
\be d\tau=3 H dt.\ee  Besides, if no additional information is available on the functional forms of the coupling and the
potential, the most natural variable to add to the former ones is the scalar field itself.

\subsection{The autonomous system}

The field equations (\ref{Raych}, \ref{Cont}, \ref{motion1}, \ref{motion2}) can be used to obtain evolution equations for
the variables (\ref{vars}) and the scalar field $\phi.$

\begin{align}  &x'=\frac{1}{2} x \left(2 y^2+z^2 \gamma \right),\label{eq0x}\\
      &y'=y^3+\frac{1}{2} \left(z^2 \gamma -2\right) y-\frac{x^2 \partial_\phi V(\phi)}{3 \sqrt{6}}+\frac{(4-3\gamma)z^2}
      {2\sqrt{6}}\partial_\phi \ln \chi (\phi),\label{eq0y}\\
      &z'=\frac{1}{2} z \left(2 y^2+\left(z^2-1\right) \gamma \right)-\frac{(4-3\gamma)y z}{2\sqrt{6}}\partial_\phi \ln
      \chi (\phi),\label{eq0z}\\
  &\phi'=\sqrt{\frac{2}{3}} y \label{eq0phi}
\end{align}
where the prime denotes derivative with respect to $\tau.$ This is
an autonomous system where the variables are subject to the
constraint \begin{eqnarray} y^2+z^2+1/3 x^2
V(\phi)=1.\label{eq22}\end{eqnarray} Observe that $y^2+z^2\leq 1$,
since $V(\phi)$ is nonnegative and the restriction (\ref{eq22})
holds.

By the \ref{Prop 4.1} in \ref{appendixA} it is possible to prove
that any combination of the sets
$$x<0,\,x=0,\,x>0,z<0,\,z=0,\,z>0$$ is an invariant set of the
flow of our dynamical system provided $\chi$ is at least of class
$C^2.$ Taking into account this result we can restrict our
attention to the flow restricted to the phase space:

\begin{equation} \Sigma=\left\{ \left(\phi,\,x,\,y,\,z\right)\in\mathbb{R}^4: x\geq 0,\,z\geq 0,\,y^2+z^2+1/3 x^2
V(\phi)=1\right\}.\label{Sigma}\end{equation}

Observe that if $x\neq 0$ and $V(\phi)>0$ we can use the constraint (\ref{eq22}) as a definition for $x.$ In this way the
evolution equation of $x$ decouples from the evolution equation of the other variables. The equation (\ref{eq0y}) can be
rewritten as

\begin{equation} y'=-\frac{(1-y^2-z^2)}{ \sqrt{6}}\partial_\phi \ln V(\phi)+\frac{(4-3\gamma)z^2}{2\sqrt{6}}\partial_\phi
\ln \chi (\phi)+y^3+\frac{1}{2} \left(z^2 \gamma -2\right) y\label{neweq0y}.\end{equation} Hence, we may then study the
dynamical system on $\mathbb{R}^3$ given by the equations (\ref{eq0x}), (\ref{neweq0y}) and (\ref{eq0z}) with the
constraint $y^2+z^2<1.$

\subsection{Main Lemma}

First we want to prove the lemma \ref{Theorem I} stating that the
orbits passing through an arbitrary point $p\in Z^+\ X^0$
interpolates between a regime where the Hubble parameter diverges
(containing an initial singularity into the past) to a regime
where the background density is negligible (into the future).
Those orbits represent cosmological solutions with non-vanishing
dimensionless background energy density and finite and positive
Hubble parameter. The result stated by the above lemma is obtained
by constructing a monotonic function defined on an invariant set
and by applying the LaSalle monotonicity principle (Theorem
\ref{theorem 4.12} in \ref{appendixA}). This lemma is very
important as a tool for investigating the past attractor or
$\omega$-limit set of the flow (see definitions
\ref{omegalimitpoint} and \ref{omegalimitset} in \ref{appendixA}):
it is necessarily located at the invariant set $\{x=0\}.$

\begin{lem}\label{Theorem I}
Let be $Z^+=\{(x,y,z,\phi)\in\Sigma: z>0\}$ and let be
$X^0=\{(x,y,z,\phi)\in\Sigma: x=0\}.$ Then for all $p\in Z^+ -
X^0$ the $\alpha$- and $\omega$-limit sets of $p$ are such that
$\alpha(p)\subset X^0$ and $\omega(p)\subset\partial Z^+,$ where
$\partial Z^+$ denotes the boundary of $Z^+.$

\end{lem}

Proof. By the proposition \ref{Prop 4.1} in \ref{appendixA} we have that
$S=Z^+ - X^0=\{(x,y,z,\phi)\in\Sigma: z>0, x>0\}$ is an invariant
set of the flow (\ref{eq0x}-\ref{eq0phi}). Let the function
$$Z(x,y,z,\phi)=\left(\frac{z}{x}\right)^2
\chi(\phi)^{-2+3\gamma/2}$$ with $\chi(\phi)>0,$ be defined on
$S.$ It is a monotonic decreasing function (in the direction of
the flow) in $S$, since its directional derivative through the
flow is $Z'=-\gamma Z$ (which is obviously negative).
\footnote{See definition \ref{Definition 4.8} in the
\ref{appendixA}.} The rank of $Z$ is $(0,\infty).$ Let be $s\in
\bar{S}-S=\partial Z^+\cup X^0.$ It is verified that
$Z(s)\rightarrow 0$ as $s\rightarrow
\partial Z^+$ and $Z(s)\rightarrow \infty$ as $s\rightarrow X^0.$
Hence, applying the LaSalle monotonicity principle (see Theorem
\ref{theorem 4.12}, in the \ref{appendixA}), we have for all $p\in
Z^+-X^0$ that $\alpha(p)\subset X^0$ and $\omega(p)\subset\partial
Z^+$, as required. $\square$

\subsection{The dynamics restricted to the invariant sets $x=0$ and $z=0.$}

Now we will to characterize the dynamics in both the invariant
sets $X^0=\{(x,y,z,\phi)\in\Sigma: x=0\}$ and
$Z^0=\{(x,y,z,\phi)\in\Sigma: z=0\}.$

The dynamics in the invariant set  $X^0$ is governed by  the differential equations \begin{eqnarray}
y'&=&\frac{1}{2} \left(1-y^2\right) \left(y(\gamma-2)+\frac{(4-3\gamma)}{\sqrt{6}}\frac{{\chi}'\left(\phi\right)}{{\chi}
\left(\phi\right)}\right),\label{eq1y}\\
\phi'&=&\sqrt{\frac{2}{3}} y, \label{eq1phi}
\end{eqnarray}
plus the algebraic equation:

\be y^2+z^2=1. \label{eq1z}\ee

The only critical point of this system (with $\phi$ bounded) is the critical point $Q$ with coordinates
$(x,\,y,\,z,\,\phi)=(0,0,1,\phi_1)$ with $\chi'(\phi_1)=0$ and $\chi(\phi_1)\neq 0.$

The critical point $Q$ represents matter dominated cosmological
solutions with the Hubble parameter diverging. Since
$\partial_{\phi}\chi(\phi_1)=0$, they are solutions with minimally
coupled scalar field (with negligible kinetic energy). The
potential function is also unimportant in the dynamics.

The eigenvalues of the liberalization around $Q$ are
$\frac{\gamma}{2},\,\gamma,\,\Delta_1\pm\sqrt{\Delta_1^2
+\Delta_2\frac{\chi''(\phi_1)}{\chi(\phi_1)}},$ where
$\Delta_1=(-2+\gamma)/{4}< 0,$ and
$\Delta_2=\left(4-3\gamma\right)/6.$

Then, the local stability of $Q$ in the invariant set $X^0$ is as
follows (we are assuming that the barotropic index $\gamma$
satisfies $0<\gamma<2$):
\begin{enumerate}
\item $Q$ is a stable focus if $0<\gamma<4/3$ and
$\chi''(\phi_1)<-\frac{\Delta_1^2\,\chi(\phi_1)}{\Delta_2} $ or
$4/3<\gamma<2$ and
$\chi''(\phi_1)>-\frac{\Delta_1^2\,\chi(\phi_1)}{\Delta_2}.$

\item $Q$ is a stable node if $0<\gamma<4/3$ and
$0>\chi''(\phi_1)\geq -\frac{\Delta_1^2\,\chi(\phi_1)}{\Delta_2}$
or $4/3<\gamma<2$ and $0<\chi''(\phi_1)\leq
-\frac{\Delta_1^2\,\chi(\phi_1)}{\Delta_2}.$ \item $Q$ is a saddle
point if  $0<\gamma<4/3$ and $\chi''(\phi_1)> 0$ or
$4/3<\gamma<2$ and $\chi''(\phi_1)< 0.$

\item $Q$ is non hyperbolic, if $\chi''(\phi_1)= 0,$ in which
case, there exists a 1-dimensional stable manifold which is
tangent to the axis $y$ at $Q$ (the associated eigenvector is
${\bf e}_y=(1,0)$). There exists also a 1-dimensional center
manifold tangent to the line $(1-\gamma/2)y-\sqrt{2/3}\phi=0$ at
$Q.$
\end{enumerate}

The dynamics in the invariant set $Z^0$ is governed by the differential equations:

\begin{eqnarray}
y'&=&\left(y^2-1\right) \left(y +\frac{\sqrt{6}}{6}\frac{ \partial_{\phi} V(\phi
   )}{V(\phi )}\right),\label{eq2y}
\end{eqnarray} and (\ref{eq1phi}), plus the equation \be y^2+1/3 x^2 V(\phi)=1, \label{eq2x}\ee  where $V(\phi)$ is given
as input.

The only critical point (with $\phi$ bounded) in the invariant set $Z^0$ is the critical point $P$ with coordinates
$x=\sqrt{\frac{3}{V(\phi_2)}},\,y=0,\,z=0,\,\phi=\phi_2$ with $\chi(\phi_2)\neq 0,\, V'(\phi_2)=0.$ The eigenvalues of
the linearization around $P$ are: $0, -\frac{\gamma}{2}, -\frac{1}{2}\pm
\frac{1}{2}\sqrt{1-\frac{4}{3}V''(\phi_2)/V(\phi_2)}.$ The zero eigenvalue has the associated eigendirection ${\bf e}_x.$

The local behavior of the critical point $P$ in the invariant set $Z^0$ is as follows:

\begin{enumerate}

\item $P$ is a saddle if  $V''(\phi_2)<0,$

\item $P$ is a stable node if $0<V''(\phi_2)\leq\frac{3}{4} V(\phi_2),$ and

\item $P$ is a stable focus if  $V''(\phi_2)>\frac{3}{4} V(\phi_2).$

\item $P$ is non hyperbolic (in the invariant set $Z^0$) if
$V''(\phi_2)=0.$

\end{enumerate}
When the orbits located at $Z^0,$ approach this critical point,
the energy density of DM and the kinetic energy density of DE
tends to zero. In this case the energy density of the Universe
will be dominated by the potential energy of DE. Hence, the
Universe would be expanding forever in a de Sitter phase.

\subsection{Conditions for the divergence of the scalar field backwards in time}

Now we will prove the Theorem \ref{Theorem IV} which, essentially,
states that if the potential and the coupling function are
sufficiently smooth functions, then for almost all the points
lying in a 4-dimensional state space, the scalar field diverges
when the orbit through $p$ is followed backwards in time. This
theorem is an extension of the theorem 1 in the reference
\cite{Foster1998sk} to STT.

\begin{thm}\label{Theorem IV}
Assume that $\chi(\phi)$ and $V(\phi)$ are positive functions of
class $C^3.$ Let $p$ be a point in $\Sigma,$ and let $O^{-}(p)$ be
the past orbit of $p$ under the flow of (\ref{eq0x}-\ref{eq0phi})
with constraint (\ref{eq22}). Then, $\phi$ is almost always
unbounded on $O^{-}(p)$ for almost all $p$.
\end{thm}

Proof. Let $p\in\Sigma$ such that $\phi$ is bounded on $O^{-}(p)$.
It follows that $O^{-}(p)$  is contained on a compact subset of
(the closure of) $\Sigma.$ Hence, that trajectory must
asymptotically approach some limit set $\alpha(p)$ (see the
analogous of theorem \ref{omegalimitsetproperties} in
\ref{appendixA} for $\alpha$ limit sets). From (\ref{eq0x}) follows
that $x$ is a monotonic increasing function through the flow and
then, it must be constant in $\alpha(p).$

There are two possibilities i) $y=z=0$ at $\alpha(p)$ or ii) $x=0$
at $\alpha(p).$

The only invariant set with $y=z=0$ (and $\phi$ bounded) is the critical  point $P$

Now we will prove that $P$ is not the past asymptote of an open set of orbits of $\Sigma.$

Observe that at least one of its associated eigenvalues has always
negative real part. Hence, by the Center Manifold Theorem
(theorem \ref{centermanifoldtheorem} in \ref{appendixA}), we can
conclude that there exists an invariant stable manifold, $E^s$, of
$P,$ intersecting $P.$ The existence of an stable manifold of
dimension $r>0$ implies that all solutions asymptotically
approaching $P$ in the past (i.e., those ones approaching $P$ as
$\tau\rightarrow -\infty$) are in an invariant unstable manifold
or a center manifold of dimension $4-r<4.$

Then, the only possibility is that $\alpha(p)$ is contained in $x=0$ (i.e., it is contained in the circumference
$y^2+z^2=1.$)

The only invariant sets for the flow given by
(\ref{eq1y}-\ref{eq1z}) are the hypersurfaces $|y|=1$ and the
critical point $Q.$  Then, there are two possibilities:
$\alpha(p)=Q$ or the $\alpha$-limit set of $p$ lies on the
hypersurface $|y|=1.$ In the last case, from (\ref{eq1phi}), we
have that $\phi$ is unbounded on $\alpha(p)$, a contradiction.

In order to complete the proof we need only to demonstrate that
$Q$ is not the past asymptote to an open set of trajectories in
$\Sigma.$

In fact, at least one of its associated eigenvalues has negative
real part. In view of the Center Manifold Theorem, we can
conclude that there exists a local stable manifold, $E^s$, that
intersects $Q$ of dimension $s>0.$ All orbits close to $Q$ in this
set exponentially approach $Q$ as $\tau\rightarrow +\infty.$ As
before, the existence of a local stable manifold of dimension
$s>0$ (equal 1 or 2) implies that all solutions past asymptotic to
$Q$ must lie on an unstable manifold or center manifold of
dimension $4-s<4.$ $\square$

Theorem \ref{Theorem IV} allow us to conclude that in order to
investigate the generic asymptotic behavior of the system
(\ref{eq0x}-\ref{eq0phi}) with restriction (\ref{eq22}) it is
necessary to study the region where $\phi=\pm\infty.$ However, as
has been investigated in \cite{Daniaetal} (where results from
\cite{Miritzis2003ym} are extended), the region $\phi=\pm\infty$
is not exclusively associated to the asymptotic behavior towards
the past. In fact, the scalar field can diverge towards the
future, provided additional requirements under the potential and
under the coupling function are fullfilled. But, this will be the purpose of a
forthcoming paper.

\subsection{The flow near $\phi=+\infty$}

In this section we will investigate the flow near $\phi=+\infty$ following the nomenclature and formalism introduced
in \cite{Foster1998sk}. Analogous results hold near $\phi=-\infty.$

By assuming  that $V,\chi\in {\cal E}^2_+,$ with exponential
orders $N$ and $M$ (the set of all class k WBI functions; see the
definitions \ref{WBI} and \ref{CkWBI} in \ref{appendixB})
respectively, we can define a dynamical system well suited to
investigate the dynamics near the initial singularity. We will
investigate the critical points therein. Particularly those
representing scaling solutions and one associated with the initial
singularity.

Let $\Sigma_\epsilon\subset \Sigma$ be the set of points in
$\Sigma$ for which $\phi>\epsilon^{-1},$ where $\epsilon$ is any
positive constant which is chosen sufficiently small so as to
avoid any points where $V$ or $\chi=0,$ thereby ensuring that
$\overline{W}_V(\vphi)$ and $\overline{W}_{\chi}(\vphi)$ are
well-defined (see definitions of $W$'s and of hatted functions in
\ref{appendixB}).

We now make the coordinate transformation

\begin{eqnarray} (x,\,y,\,z,\,\phi) \stackrel{\vphi=f(\phi)}{\longrightarrow} (x,\,y,\,z,\,\vphi)\label{carta2}\end
{eqnarray} on $\Sigma_\epsilon,$ where $f(\phi)$ tends to zero as
$\phi$ tends to $+\infty$ and has been chosen so that the
conditions i)-iii) of definition \ref{CkWBI} in \ref{appendixB}
are satisfied with $k=2.$

Substituting this new coordinates in the equations (\ref{eq0x}), (\ref{neweq0y}) and (\ref{eq0z}) we obtain the
3-dimensional dynamical system:

\begin{align}
&y'=y^3+\frac{1}{2} \left(z^2 \gamma -2\right) y-\frac{(1-y^2-z^2)}{ \sqrt{6}}\left(\overline{W}_V+N\right)+\frac{z^2
(4-3
   \gamma)}{2 \sqrt{6}}\left(\overline{W}_{\chi}+M\right),\label{eqinfy}\\
&z'=\frac{1}{2} z \left(2 y^2+\left(z^2-1\right) \gamma \right)+\frac{y z (-4+3 \gamma )}{2 \sqrt{6}}\left(\overline{W}_
{\chi}+M\right),\label{eqinfz}\\
&\vphi'=\sqrt{\frac{2}{3}} \overline{f'} y. \label{eqinfphi}
  \end{align}

We may identify $\Sigma_\epsilon$ with its projection into $\mathbb{R}^3$ so that we have
$\Sigma_\epsilon=\left\{0<\varphi< f(\epsilon^{-1}),\, 0\leq y^2+z^2<1\right\}.$ The variable $x$ can be treated as a
function on $\Sigma_\epsilon$ defined by the constraint equation which becomes

\be y^2+z^2+1/3 x^2 \overline{V}(\vphi)=1\label{eq35}.\ee The directional derivative of $x$ along the flow generated by
(\ref{eqinfy}-\ref{eqinfphi}) may be obtained directly by equation (\ref{eq0x}).

Since $\overline{f'},$ $\overline{W}_V$ and $\overline{W}_{\chi}$ are $C^2$ at $\vphi=0$ we may extend
(\ref{eqinfy}-\ref{eqinfphi}) onto the boundary of $\Sigma_\epsilon$ to obtain a $C^2$ system on the closure of
$\Sigma_\epsilon$, i.e., $\overline{\Sigma_\epsilon}.$
From definition \ref{CkWBI}, $\overline{f'},$ $\overline{W}_V$ and $\overline{W}_{\chi}$ vanish at the origin and are
each of second order or higher in $\vphi$ and $\overline{f'}$ is negative on $\Sigma_\epsilon.$

\subsubsection{Critical points.}\label{criticalpoints}

The system (\ref{eqinfy}-\ref{eqinfphi}) admits the critical
points labelled by $P_i$, $i\in\{1,2,3,4,5,6\}.$ In the following
we discuss the existence and the stability conditions for the
critical points. In the table \ref{crit} are displayed the values
of some cosmological magnitudes of interest for the critical
points (the deceleration parameter, the effective equation of
stare (EoS) parameter for the total matter, etc).

\begin{enumerate}

\item The critical point $P_1$ with coordinates $y=-1,$ $z=0$ and
$\vphi=0$ exists for all the values of the free parameters. The
eigenvalues of the linearized system around $P_1$ are
$\lambda_{1,1}=2-\sqrt{2/3} N,\,\lambda_{1,2}=
\frac{2-\gamma}{2}-\frac{M(-4+3\gamma)}{2\sqrt{6}}$ and
$\lambda_{1,3}=0.$ Hence the critical point is non hyperbolic,
then the Hartman-Grobman theorem does not apply. By the Center
Manifold Theorem there exists:
\begin{enumerate}

\item an stable invariant subspace of dimension two (tangent to
the y-z plane) if: i) the potential is a WBI function (see definition \ref{WBI} in \ref{appendixB}) of
exponential order $ N>\sqrt{6}$ and the coupling function is a WBI
function of exponential order $M<-\frac{\sqrt{6} (\gamma
   -2)}{3 \gamma -4}$ (provided $0<\gamma <\frac{4}{3}$), or  ii) the barotropic index satisfies $\frac{4}{3}<\gamma <2$,
   the potential is a WBI function of exponential order $N>\sqrt{6}$ and the coupling function is a WBI function of
   exponential order $M>-\frac{\sqrt{6} (\gamma  -2)}{3 \gamma -4};$

\item an unstable invariant subspace of dimension two (tangent to the y-z plane) provided the potential is a WBI function
of exponential order $N<\sqrt{6}$ and the coupling function is a WBI function of exponential order $M$ such that $M>-
\frac{\sqrt{6} (\gamma
   -2)}{3 \gamma -4}$ (respectively, $M<-\frac{\sqrt{6} (\gamma   -2)}{3 \gamma -4}$) provided $0<\gamma <\frac{4}{3}$
   (respectively, $\frac{4}{3}<\gamma <2$);

\item a 1-dimensional center manifold which is tangent to the
critical point in the direction of the axis $\vphi$. This center
manifold can be 2-dimensional or even 3-dimensional (see the
discussion on the point 3).

\end{enumerate}

\item The critical point $P_2$ with coordinates $y=1,$ $z=0$ and
$\vphi=0$ exists for all the values of the free parameters. The
eigenvalues of the linearized system around $P_2$ are
$\lambda_{2,1}=2+\sqrt{2/3} N,$ and $\lambda_{2,2}=\lambda_{1,2}$
and $\lambda_{2,3}=0$ (see point 1). Hence the critical point is
non hyperbolic, then the Hartman-Grobman theorem does not apply.
By the Center Manifold Theorem there exists:
\begin{enumerate}

\item an stable invariant subspace of dimension two (tangent to the y-z plane) if: i) $ N<-\sqrt{6},$  $M>\frac{\sqrt{6}
(\gamma
   -2)}{3 \gamma -4}$ for $0<\gamma <\frac{4}{3}$, or  ii)  $\frac{4}{3}<\gamma <2,$  $N<-\sqrt{6}$ and  $M<\frac{\sqrt{6}
    (\gamma  -2)}{3 \gamma -4};$

\item an unstable invariant subspace of dimension two (tangent to the y-z plane) provided $N>-\sqrt{6},$ and $M$ such that
 $M<\frac{\sqrt{6} (\gamma
   -2)}{3 \gamma -4}$ (respectively  $M>\frac{\sqrt{6} (\gamma   -2)}{3 \gamma -4}$) provided $0<\gamma <\frac{4}{3}$
   (respectively $\frac{4}{3}<\gamma <2$);

\item a 1-dimensional center manifold which is tangent to the
critical point in the direction of the axis $\vphi$. This center
manifold can be 2-dimensional or even 3-dimensional (see the
discussion on the point 3).

\end{enumerate}

In the following section we shall study the initial spacetime (big bang) singularity. The critical points $P_{1,2}$ can
account for that singularity. They are in the same phase portrait for the values $-\sqrt{6}<N<\sqrt{6}$ and $-\frac{\sqrt
{6} (\gamma   -2)}{3 \gamma -4}<M<\frac{\sqrt{6} (\gamma   -2)}{3 \gamma -4}$ and $0<\gamma<\frac{4}{3}$ (in which case
they have a 2-dimensional unstable manifold and a 1-dimensional center respectively). It is easy to show that the Hubble
parameter (and the matter density) of the cosmological solutions associated to these points diverges into the past. The
scalar field also diverges, it equals to $+\infty$ (respectively $-\infty$) for $P_1$ (respectively $P_2$). However, even
in this case, the past attractor corresponds to $P_1$ since $\overline{f'}<0$ and for $y>0$ the orbits enter the phase
portrait and $P_2$ acts as a saddle. The last point can be a past attractor only on a set of measure zero (if $\vphi=0$).

\item The critical point $P_3$  with coordinates $y= \frac{M (-4+3 \gamma )}{\sqrt{6} (-2+\gamma )},$ $z=\sqrt{1-\frac{M^2
 (4-3 \gamma )^2}{6(-2+\gamma )^2}}$ and $\vphi=0$ exists if $0<\gamma <\frac{4}{3}$ and $-\frac{\sqrt{6} (-2+\gamma )}
 {-4+3 \gamma }\leq M\leq \frac{\sqrt{6} (-2+\gamma )}{-4+3 \gamma }.$ The eigenvalues of the matrix of derivatives
 evaluated at the critical point are $\lambda_{3,1}=\frac{6 (\gamma -2)^2-M^2 (4-3 \gamma )^2}{12 (\gamma -2)},\; \lambda_
 {3,2}=-\frac{3 \gamma  M^2}{2}+(M+N) M+\frac{2 (N-M) M}{3 (\gamma -2)}+\gamma,$ and $\lambda_{3,3}=0.$ Hence the critical
  point is non hyperbolic, then the Hartman-Grobman theorem does not apply.  Under the above existence conditions we find, by the Center Manifold Theorem, that there exists a stable manifold of dimension two for the values of the parameters:
  i) $M<0$ and $N>\frac{M^2 (4-3 \gamma )^2-6 (\gamma -2) \gamma }{2 M (3 \gamma -4)}$ or ii) $M>0$ and $N<\frac{M^2 (4-3
  \gamma )^2-6 (\gamma -2) \gamma }{2 M (3 \gamma -4)}$. Otherwise there exists an unstable manifold of dimension one
  (in this case the stable subspace is 1-dimensional). The center manifold is in both cases 1-dimensional. If $M=\mp\frac
  {\sqrt{6} (-2+\gamma )}{-4+3 \gamma }$ this critical point reduces to $P_{1,2}.$ In this case the center subspace is
  2-dimensional and is spanned by the eigenvectors ${\bf e}_z=\left(
\begin{array}{l}
 0 \\
 1 \\
 0
\end{array}
\right),\; {\bf e}_\vphi=\left(
\begin{array}{l}
 0 \\
 0 \\
 1
\end{array}
\right).$ The center manifold is tangent to the center subspace at the critical point. If additionally $\left|N\right|=
\sqrt{6},$ the center manifold is 3-dimensional.

\item The critical point $P_4$ with coordinates
$y=-\frac{N}{\sqrt{6}},\, z=0$ and $\vphi=0$ exists if
$\left|N\right|\leq\sqrt{6}.$ Observe that this point reduces to
$P_{1,2}$ if $N^2=6.$ The matrix of derivatives evaluated at the
critical point has the eigenvalues $\lambda_{4,1}=\frac{1}{6}
\left(N^2-6\right)\leq 0,$ $\lambda_{4,2}=\frac{1}{6} N (2
M+N)-\frac{1}{4} (M N+2) \gamma$ and $\lambda_{4,3}=0.$ Hence the
critical point is non hyperbolic and, as before, the
Hartman-Grobman theorem does not apply. However, we can use the
Center Manifold Theorem to investigate the stability of this
critical point. The structure of the center manifold is as
follows:

\begin{enumerate}

 \item if $\lambda_{4,1}<0$ and $\lambda_{4,2}\neq 0$ the center manifold is spanned by ${\bf e}_\vphi.$ Then, it is
 1-dimensional. Before analyzing this case in detail, we will provide additional information about the structure of the
 center manifold.

 \item if $M=\frac{2 \left(N^2-3 \gamma \right)}{N (3 \gamma -4)}$ and $N^2<6$, the center subspace is spanned by the
 eigenvectors ${\bf e}_z$ and ${\bf e}_\vphi.$
 \item if $N^2=6$ and $M\neq\mp\frac{\sqrt{6} (-2+\gamma )}{-4+3 \gamma },$ it is spanned by the eigenvectors
 ${\bf e}_y=\left(
\begin{array}{l}
 1 \\
 0 \\
 0
\end{array}
\right),$ and ${\bf e}_\vphi.$
\item  if $N^2=6$ and $M=\mp\frac{\sqrt{6} (-2+\gamma )}{-4+3 \gamma },$ the center manifold is 3-dimensional.

\end{enumerate}

The local behavior described in the cases above (excluding the first case) is in some way special. It requires fine tuning
 of the free parameter. However, the typical behavior (in the invariant set $z=0$) is the existence of a one dimensional
 center manifold $C_N$ through $P_4$, which is tangent to the $z$-axis (if $\lambda_{4,1}<0$ and $\lambda_{4,2}\neq 0$).
 $C_N$ is an exponential attractor on a sufficiently small neighborhood of $P_4.$ It is intuitively obvious from the
 geometry (for instance, observe figure 1), that any solutions past asymptotic to $P_4$ must lie on the center manifold.

Let us investigate the case in which $\lambda_{4,1}<0$ and $\lambda_{4,2}\neq 0.$ Of course, in this case the stable
manifold is at least 1-dimensional (and as we mentioned before the center manifold is 1-dimensional).

The structure of the stable subspace is as follows:

\begin{enumerate}

\item if the potential is of exponential order zero ($N=0$), then, the critical point has coordinates $(0,0,0).$ The
eigenvalues of the linearization are $\left(-1,0,-\frac{\gamma}{2}\right)$ and in this case, the stable subspace is
generated by the eigenvectors ${\bf e}_y, {\bf e}_z;$

\item if $0<\gamma<\frac{4}{3},$ $-\sqrt{6}<N<0,$ and $M>\frac{2 \left(N^2-3 \gamma \right)}{N (3 \gamma -4)};$ or

\item if $\frac{4}{3}<\gamma<2,$ $-\sqrt{6}<N<0,$ and $M<\frac{2 \left(N^2-3 \gamma \right)}{N (3 \gamma -4)};$ or

\item if $0<\gamma<\frac{4}{3},$ $0<N<\frac{4}{3},$ and $M<\frac{2 \left(N^2-3 \gamma \right)}{N (3 \gamma -4)};$ or

\item if $\frac{4}{3}<\gamma<2,$ $0<N<\sqrt{6},$ and $M>\frac{2 \left(N^2-3 \gamma \right)}{N (3 \gamma -4)}$ the stable
subspace is generated by the eigenvectors ${\bf e}_y, {\bf e}_z.$

\item By interchanging $>$ and $<$ in the inequalities for $M$ in
the cases (b)-(e) we find that the stable manifold is
1-dimensional and is tangent to the critical point in the
direction of ${\bf e}_y$ (accordingly, the unstable subspace is
spanned by ${\bf e}_z$).

\end{enumerate}

\item The critical points $P_{5,6}$ with coordinates $y=\frac{\sqrt{6} \gamma }{M (3 \gamma -4)-2 N},\; z=\mp\frac{\sqrt{4
 N (2
   M+N)-6 (M N+2) \gamma }}{2 N+M (4-3 \gamma )}$  (respectively) exists if the following conditions are simultaneously
   satisfied: $4 N (2 M+N)-6 (M N+2) \gamma \geq 0,$  $\mp\left({2 N+M (4-3 \gamma )}\right)> 0$ and $\frac{4 N^2+M (8-6
   \gamma ) N+6 (\gamma -2) \gamma }{(2 N+M (4-3 \gamma
   ))^2}\leq 1$ (i.e., the critical points are real-valued, and they are inside the cylinder $\overline{\Sigma_\epsilon}$).

The associated eigenvalues are

 $$\lambda_{5,6}^\pm=\frac{\alpha}{\beta}\pm\frac{\sqrt{8 \left(\beta^2+27 \gamma ^2\right) \alpha^2-2 \beta (\gamma
   -4) \left(\beta^2-216 \gamma ^2\right) \alpha-(\gamma -2) \left(\beta^2-216 \gamma
   ^2\right)^2}}{6 \sqrt{6} \beta \gamma }$$
and $\lambda_{5,6}=0,$ where
$\alpha=3\left(N(\gamma-2)+M(3\gamma-4)\right)$ and
$\beta=2\left(2 N-M(3\gamma-4)\right).$ Assuming that the
conditions for existence are satisfied, we can analyze the
stability of the critical points by means of the Center Manifold
Theorem. We find that the non null eigenvalues can not be either
complex conjugates with positive real parts or real-valued with
different sign, then the unstable subspace of $P_{5,6}$ is the
empty set. Then, the stable subspace is 2-dimensional (provided
$\lambda_{5,6}$ is the only null eigenvalue). When the orbits are
restricted to this invariant set, the point $P_{5,6}$ acts as an
stable spiral (if the eigenvalues are complex conjugated) or as a
node (if the eigenvalues are negative reals). The conditions on
the parameters for those cases are very complicated to display
here.

If $M=\frac{2 \left(N^2-3 \gamma \right)}{N (3 \gamma -4)}$ the $P_{5,6}$ reduces to $P_4$ but in this case, the center
manifold is 2-dimensional and is spanned by ${\bf e}_z,\,{\bf e}_\vphi.$

\end{enumerate}

\begin{table*}[htbp]
\begin{center}
\begin{tabular}
{|@{\hspace{0.03in}}c@{\hspace{0.03in}}|@{\hspace{0.03in}}c
@{\hspace{0.03in}}|
@{\hspace{0.03in}}c@{\hspace{0.03in}}|@{\hspace{0.03in}}c
@{\hspace{0.03in}}| @{\hspace{0.03in}}c@{\hspace{0.03in}}|
@{\hspace{0.03in}}c@{\hspace{0.03in}}|@{\hspace{0.03in}}c @{\hspace{0.03in}}c@{\hspace{0.05in}}|} \hline \small\small
Point& $y$ & $z$  & $\Omega_{de}$  &
$w_{\text{tot}}$ & Acceleration?
\\
[0.1cm] \hline
\hline &&&&&\\[-0.3cm]
$P_1$ &-1 &0
& 1&
 1 &  no \\[0.2cm]
\hline&&&&&\\[-0.3cm]
$P_2$ &1 &0  &
1 & 1 &  no \\
[0.2cm]
\hline&&&&&\\[-0.3cm]
$P_3$ &$\delta$ &$\displaystyle{\sqrt{1-\delta^2}}$   &
$\delta^2$ & $\displaystyle{\gamma+(\gamma-1)\delta}$  & $\displaystyle{0<\gamma<\frac{2}{3}}$ and \\&&&&& $\displaystyle{\left|M\right|<\Gamma}$\\[0.2cm]
\hline&&&&&\\[-0.3cm]
$P_4$ &$-\frac{N}{\sqrt{6}}$ & 0  &
1 & $-1+\frac{N^2}{3}$ &  $N^2<2$\\
[0.2cm]
\hline&&&&&\\[-0.3cm]
$P_{5,6}$ &$-\frac{6 \sqrt{6} \gamma }{\beta }$&$\mp\frac{\sqrt{\frac{2 \beta  (2 \alpha +\beta )}{\gamma }-432 \gamma }}{\beta
   }$& $-\frac{2 (2 \alpha +\beta )}{\beta  \gamma }+\frac{432 \gamma }{\beta ^2}+1$ & $\frac{(\gamma +2) \beta ^2+4 \alpha  (\gamma +1) \beta -432 \gamma
   ^2}{\beta ^2 \gamma }$& $\frac{\alpha}{\beta}<-\frac{1}{3}$\\[0.2cm]
\hline
\end{tabular}
\end{center}
\caption[crit]{\label{crit} The properties of the critical points
for the system (\ref{eqinfy}-\ref{eqinfphi}). We use the notations $\alpha=3\left(N(\gamma-2)+M(3\gamma-4)\right),$  $\beta=2\left(2 N-M(3\gamma-4)\right),$  $\delta=\displaystyle\frac{M(3\gamma-4)}{\sqrt{6}(\gamma-2)},$ and $\Gamma=\frac{\sqrt{2(\gamma-2)(3\gamma-2)}}{4-3\gamma}.$}
\end{table*}

\subsection{The flow near $\phi=-\infty$}\label{minusinfinity}

With the purpose of complementing the global analysis of the
system it is necessary investigate its behavior near
$\phi=-\infty.$ It is an easy task since the system (\ref{Raych},
\ref{Cont}, \ref{motion1}, \ref{motion2}) is invariant under the
transformation of coordinates

$$(\phi, \dot\phi)\rightarrow -(\phi, \dot\phi),\; V\rightarrow U,\; \chi\rightarrow \Xi,$$ where $U(\phi)=V(-\phi)$ and
$\Xi(\phi)=\chi(-\phi).$ Hence, for a particular potential $V,$
and a particular coupling function $\chi$, the behavior of the
solutions of the equations (\ref{Raych}, \ref{Cont},
\ref{motion1}, \ref{motion2}) around $\phi=-\infty$ is equivalent
(except for the sign of $\phi$) to the behavior of the system near
$\phi=\infty$ with potential and coupling functions $U$ and $\Xi,$
respectively.

If $U$ and $\Xi$ are of class $\mathcal{E}^2_+,$ the preceding  analysis in $\bar{\Sigma}_\epsilon$ can be applied
(with and adequate choice of $\epsilon$).

In the following we will denote $\mathcal{E}^k$ to the set of
class $C^k$ functions well behaved in both $+\infty$ and
$-\infty.$ We will use Latin uppercase letters with subscripts
$+\infty$ and $-\infty,$ respectively to indicate the exponential
order of $\mathcal{E}^k$ functions in $+\infty$ and in $-\infty.$

\subsection{The global geometric structure of the phase space}\label{estructura}

Let be $\Omega(x_0)$ the region of the phase space given by  (\ref{Sigma}) with $x<x_0,$ then, since $x$ is monotonic
decreasing, this set equals the union of its past orbits.

The procedure to define a coordinate system near $-\infty,$ given
in the section \ref{minusinfinity},  can be used to embed
$\Omega(x_0)$ as a compact differentiable 4-dimensional manifold
$\Sigma(x_0)$ such that the vector field defined by
(\ref{eq0x}-\ref{eq0phi}) can be smoothly extended over
$\Sigma(x_0).$

With this purpose we define an atlas as follows.

First, the interior of $\Sigma(x_0)$ is defined as the set $$\left\{(\phi,x,y,z)\in\mathbb{R}^4: 0<x<x_0, y^2+z^2<1\right\}
$$ where we use the local chart (coordinate system) given by (\ref{vars}). Obviously, this set is bounded in the variables
 $x,\,y,\,z.$ We define a second local chart $(\varphi,x,y,z)$ in the open subset of this set for which $x>\epsilon^{-1},$
  for $\epsilon$ small enough, by (\ref{carta2}). An analogous local chart $(\varphi,x,y,z)$ can be defined near
  $\phi=-\infty$ by the procedure given in the section \ref{minusinfinity}.

The construction is completed by attaching a boundary which is defined taking the union of $x=0,\, x=x_0,\, \varphi=0$ and
 the circumference $y^2+z^2=1$ to each local chart.

By construction, $\Sigma(x_0)$ is compact and it is embedded in $\mathbb{R}^4.$

The vector field defined by (\ref{eq0x}-\ref{eq0phi}) can be smoothly extended over the boundary of $\Sigma(x_0)$ such
that $\Sigma(x_0)$ is the union of its past orbits. $\Omega(x_0)$ is a 3-dimensional hypersurface embedded in $\Sigma(x_0).
$ It is important to note that $\Omega(x_0)$ approaches the non-physical boundary along the intersection of the plane
$x=0$ with the plane $\varphi=0$ and the circumference $y^2+z^2=1.$ This set is called non-physical boundary of
$\Omega$ and denoted by $\partial \Omega.$

\section{The initial space-time singularity}

In this section we will study the initial space-time (Big-Bang)
singularity. The critical points $P_{1,2}$ can represent such a
singularity. They live at the same phase space for the values of
$M,$ $N$ and $\gamma$ in the intervals $-\sqrt{6}<N<\sqrt{6},$
$-\frac{\sqrt{6} (\gamma   -2)}{3 \gamma -4}<M<\frac{\sqrt{6}
(\gamma   -2)}{3 \gamma -4}$ and $0<\gamma<\frac{4}{3}$ (in this
case, they have an unstable 2-dimensional manifold and a center
1-dimensional manifold). It is easy to show that the Hubble
parameter and the matter energy density of the associated
cosmological solutions diverge towards the past. The scalar field
diverges too, and it is equal to $+\infty$ and to $-\infty$ for
$P_1$ and $P_2$ respectively. However, even in this case, the
possible past attractor corresponds to $P_1$ since
$\overline{f'}<0$ whereas for  $y>0$ the orbits enter the phase
space and $P_2$ acts as a saddle. The critical point $P_2$ can act
as a past attractor only in a set of initial conditions of measure
zero (when $\varphi=0$).

\subsection {Analysis near $P_1.$}

From the analysis in section \ref{criticalpoints}, it seems
reasonable to think that the initial space-time singularity can be
associated to the critical point $P_1.$ Its unstable manifold is
2-dimensional provided $N<\sqrt{6}.$ The asymptotic behavior of
neighboring solutions to $P_1$ can be approximated, for $\tau$
negative large enough, as \be
y(\tau)=-1+O(e^{\lambda_{1,1}\tau}),\,
z(\tau)=O(e^{\lambda_{1,2}\tau})\label{yzapprox}.\ee  By
substitution of (\ref{yzapprox}) in (\ref{eq0phi}), and
integrating the resulting equation, we obtain
\be\phi(\tau)=\sqrt{\frac{2}{3}}\left(-\tau+\tilde{\phi}\right)+O(e^{\lambda_{1,1}\tau})\label{phiapprox}.\ee
Then, by expanding around $\tau=-\infty$ up to first order, we get
$$\varphi=f\left(\sqrt{\frac{2}{3}}\left(-\tau+\tilde{\phi}\right)+O(\frac{1}{\tau})^2\right)+O(e^{\lambda_{1,1}\tau})=
f\left(\sqrt{\frac{2}{3}}\left(-\tau+\tilde{\phi}\right)\right)+O(e^{\lambda_{1,1}\tau})+h,$$
where $h$ denotes higher order terms to be discarded.

Then we have a first order solution to (\ref{eqinfy}-\ref{eqinfphi}). Also, by substitution of (\ref{yzapprox}) in
(\ref{eq0x}) and solving the resulting differential equation with initial condition $x(0)=x_0$ we get the first order
solution \be x=x_0 e^\tau.\label{xfirstorder}\ee Then, we have $t-t_i=\frac{1}{3}\int x(\tau) d\tau=1/3 x_0 e^\tau.$
For simplicity let us set $t_i=0.$

Neglecting the error terms, we have the following expressions
\begin{eqnarray}
H&=&x^{-1}=\left(x_0 e^\tau\right)^{-1}=\frac{1}{3 t},\;
\phi=\sqrt{\frac{2}{3}}\left(-\tau+\tilde{\phi}\right)=-\sqrt{\frac{2}{3}}\ln\frac{ t}{c},\nonumber\\
\dot\phi&=&-\sqrt{\frac{2}{3}}
t^{-1},\;\rho=0,\label{massless}\end{eqnarray} where
$c=1/3x_0e^{\tilde{\phi}}.$ This asymptotic solution corresponds
to the exact solution of (\ref{Raych}, \ref{Cont}, \ref{motion1},
\ref{motion2}) when V vanishes identically and $\chi$ is a
constant (the minimal coupling case). Hence, there exists a
generic class of massless minimally coupled scalar field
cosmologies in a vicinity of the initial space-time singularity.

The above idea can be stated, more precisely, as the

\begin{thm}\label{initialsingularity}
Let be $V\in\mathcal{E}_+^2$ with exponential order $N$ satisfying
$N<\sqrt{6}$ and let be $\chi\in\mathcal{E}_+^2$ with exponential
order $M$ such that
\begin{enumerate}
\item[i)] $0<\gamma<\frac{4}{3}$ and $M>-\frac{\sqrt{6}(\gamma-2)}{3\gamma-4}$ or
\item[ii)] $\frac{4}{3}<\gamma<2$ and $M<-\frac{\sqrt{6}(\gamma-2)}{3\gamma-4}$
\end{enumerate}
Then, there exist a neighborhood $\mathcal{N}(P_1)$ of $P_1$ such
that for each $p\in \mathcal{N}(P_1)$ the orbit $\psi_p$ past
asymptotic to $P_1$ and the associated cosmological solution is:
\begin{align}
&H= \frac{1}{3t} +O\left(\epsilon_V(t)\right),\\
&\phi= -\sqrt{\frac{2}{3}}\ln \frac{t}{\tilde{c}} + O\left(t\epsilon_V(t)\right),\\
&\dot\phi=-\sqrt{\frac{2}{3}} t^{-1}+ O\left(\epsilon_V(t)\right),
\\&\rho=\chi_0^2 t^{-\gamma}\chi\left(-\sqrt{\frac{2}{3}}\ln \frac{t}{\tilde{c}}\right)^{\frac{3\gamma}{2}-2}\left(1+ O
\left(t\epsilon_V(t)\right)\right)\end{align}
where $\epsilon_V(t)=t V\left(-\sqrt{\frac{2}{3}}\ln \frac{t}{\tilde{c}}\right).$
\end{thm}

Before proceeding to the proof of this theorem, let us make a few
comments.

Since $V\in \mathcal{E}_+^2$ has exponential order $N$, then, by applying Theorem 2 in \cite{Foster1998sk} we have
$$\lim_{t\rightarrow 0} t^\alpha V(-\sqrt{\frac{2}{3}}\ln{\frac{t}{c}})=\lim_{\phi\rightarrow \infty} e^{-\sqrt{\frac{3}
{2}}\alpha\phi}V(\phi)=0,\;\forall \alpha>\sqrt{\frac{2}{3}}N.$$

Then, for $N<\sqrt{6}$ the error terms $O\left(\epsilon_V(t)\right)$ and $O\left(t \epsilon_V(t)\right)$ are dominated by
the first order terms. If $N<\sqrt{\frac{3}{2}}$ both error terms tend uniformly to zero.

On the other hand, since $\chi\in \mathcal{E}_+^2$ has exponential order $M,$ then,  $t^{-\gamma}\chi(-\sqrt{\frac{2}{3}}\ln{\frac{t}{c}})^{\frac{3\gamma}{2}-2}$ tends uniformly to zero, as $t\rightarrow 0$ in the cases i) $0<\gamma<\frac{4}{3}
$ and $M>-\frac{\sqrt{6}\gamma}{3\gamma-4}$ or ii) $\frac{4}{3}<\gamma<2$ and $M<-\frac{\sqrt{6}\gamma}{3\gamma-4}.$
In the above cases, the matter tends uniformly to zero as $P_1$ is approached.

\paragraph{\bf Proof of Theorem \ref{initialsingularity}.} From the equation (\ref{eq0x}) and using (\ref{eq35}) as a
definition for $y$ we get the equation:
$$\frac{d\ln x}{d\tau}=\left(\frac{\gamma}{2}-1\right)z^2+1-\frac{1}{3} x^2\overline{V}(\vphi).$$ Using the first order
expressions $z=O(e^{\lambda_{1,2}\tau})$ and $x=x_0e^\tau$ we have
the differential equation for $x$:

\be \frac{d\ln x}{d\tau}=1-\frac{1}{3} x_0^2 \overline{V}(\vphi)e^{2\tau}+h\label{logx}.\ee Where we denote by $h$ any
collection of higher order terms to be discarded.

By integrating in both sides of (\ref{logx}) we find the solution

\begin{eqnarray}
x&=&x_0 e^\tau \exp\left(-\frac{x_0^2}{3\lambda_{1,1}}\overline{V}(\vphi) e^{2\tau}\right)+h\nonumber\\
&=&x_0 e^\tau\left(1-\frac{x_0^2}{3\lambda_{1,1}}\overline{V}(\vphi)e^{2\tau}\right)+h.\label{xsecondorder}
\end{eqnarray}

In the above deduction we have used the auxiliary result proved in \cite{Foster1998sk}:

\be\int \overline{V}(\vphi) e^{2\tau}d\tau=\frac{\overline{V}(\vphi) e^{2\tau}}{\lambda_{1,1}}+h\label{auxiliar},\ee
which is valid if $V\in{\cal E}_+^2$ with exponential order $N<\sqrt{6}.$ We have used, also, the approximation
$e^u\approx 1+u.$

Now we want to derive an second order expression for $t$.

\begin{eqnarray}
t&=& \frac{1}{3}\int x(\tau) d\tau=\frac{1}{3}\int \left(x_0 e^\tau\left(1-\frac{x_0^2}{3\lambda_{1,1}}\overline{V}(\vphi)
e^{2\tau}\right)\right)d\tau+h\nonumber\\
&=& \frac{1}{3} x_0 e^\tau \left(1-\frac{x_0^2}{9\lambda_{1,1}}\overline{V}(\vphi) e^{2\tau}\right)+h.\label{tsecondorder}
\end{eqnarray}
In the above deduction we have used the auxiliary result proved in
\cite{Foster1998sk}: \be \int \overline{V}(\vphi)
e^{3\tau}d\tau=\frac{1}{3}\overline{V}(\vphi)
e^{3\tau}+h\label{estimation1}\ee

The equation (\ref{tsecondorder}) may be inverted, to second order, to give

$$x_0 e^{\tau}=3\left(t+\frac{\overline{V}(\vphi)}{\lambda_{1,1}}t^3\right)+h.$$ Substituting this result in
(\ref{xsecondorder}) we get:

\be x(t)=3 t-\frac{6 V(\phi) t^3}{\lambda_{1,1}}+h.\label{xsecondorderent}\ee and then,
\be H(t)=\frac{1}{x(t)}=\frac{1}{3 t}+\frac{2 {V}(\phi) t}{3 {\lambda_{1,1}}}+h.\label{expansionH}\ee

The equation (\ref{eqinfy}) can be written as \be \frac{d\ln y}{d\tau}= \left(-1+\frac{\gamma}{2}\right)z^2-\frac{1}{3}x^2
\overline{V}(\vphi)-\frac{(1-y^2-z^2)}{ \sqrt{6} y}\left(\overline{W}_V+N\right)+\frac{z^2 (4-3
   \gamma)}{2 \sqrt{6}y}\left(\overline{W}_{\chi}+M\right).\nonumber\ee

By the same arguments as in the deductions of (\ref{xsecondorder}) and (\ref{xsecondorderent}) we get

\begin{eqnarray} y&=&-1+\frac{x_0^2}{3\lambda_{1,1}} e^{2\tau}\overline{V}(\vphi)+h\nonumber\\
&=&-1+\frac{3 {V}(\phi) {t}^2}{{\lambda_{1,1}}}+h\label{ysecondorderent}.\end{eqnarray}

Combining the expansions (\ref{ysecondorderent}) and (\ref{xsecondorderent}) in $\dot\phi(t)=\frac{\sqrt{6} y}{x},$ we
find

\be\dot\phi(t)=-\sqrt{\frac{2}{3}} \left(\frac{1}{t}-\frac{t
 {V}(\phi)}{\lambda_{1,1} }\right)+h.\label{expansiondotphi}\ee This equation can be integrated up to second order to get

\be \phi(t)=-\sqrt{\frac{2}{3}}\left(\ln \frac{t}{\tilde{c}}-\frac{{V}(\phi) {t}^2}{2\lambda_{1,1}}\right)+h.\label
{expansionphi}\ee

The equation (\ref{eqinfz}) can be written as

$$\frac{d \ln z}{d\tau}=-\frac{1}{3} x^2\overline{V}(\vphi)+(1-\frac{\gamma}{2})(1-z^2)+\frac{y(-4+3\gamma)}{2\sqrt{6}}
\overline{W}_V(\vphi),$$ where we have used the constraint
equation (\ref{eq35}) as a definition for $y^2.$

By the same arguments as in the deductions of (\ref{xsecondorder}) and (\ref{xsecondorderent}) we get

\begin{equation} z=z_0\exp\left(\lambda_{1,2}\tau+\frac{4-3\gamma}{2\sqrt{6}}\int \overline{W}_{\chi}(\vphi)d\tau\right)
\left(1-\frac{x_0^2}{3\lambda_{1,1}}\overline{V}(\vphi)e^{2\tau}\right)+h.\label{zexpression}\end{equation}

By definition $$\overline{W}_{\chi}(\vphi)=\frac{\chi'(f^{-1}(\vphi))}{\chi(f^{-1}(\vphi))}-M.$$ Then, by using the first
 order expression $$f^{-1}(\vphi)=\phi=\sqrt{\frac{2}{3}}(-\tau+\phi_0)+O(e^{\lambda_{1,1}\tau})+h$$ (as derived in former
  sections) and integrating out the resulting expression with respect to $\tau$ we get the estimation

\be\int \overline{W}_{\chi}(\vphi) d\tau=-\sqrt{\frac{3}{2}}\ln \overline{\chi}(\vphi)d\tau-M\tau+h.\label{intWv}\ee

By substitution of (\ref{intWv}) in (\ref{zexpression}) we get
\begin{eqnarray}
z&=&z_0 e^{(1-\frac{\gamma}{2})\tau}\overline{\chi}(\vphi)^{-1+\frac{3\gamma}{4}}\left(1-\frac{x_0^2}{3\lambda_{1,1}}
\overline{V}(\vphi)e^{2\tau}\right)+h\nonumber\\
&=&\chi_0 {t}^{1-\frac{\gamma}{2}} {{\chi}}(\phi)^{\frac{3 \gamma }{4}-1}+h\label{zsecondorderent}
\end{eqnarray} where $\chi_0=z_0\left(\frac{x_0}{3}\right)^{-1+\frac{\gamma}{2}}.$

Combining the expansions for $z$ and $x$ in $\rho=\frac{3z^2}{x}$ we find

\be\rho =\frac{1}{3}{\left(1+\frac{4}{\lambda_{1,1}} {V}(\phi) {t}^2\right) {{\chi}}{(\phi)}^{\frac{3 \gamma
   }{2}-2} {\chi_0}^2 {t}^{-\gamma }}+h.\label{expansionrho}\ee

Observe that the second term, $h,$ on the right-hand of
(\ref{expansionphi}) tends to zero when $t\rightarrow 0.$ This
allows to Taylor expand $V$ and $\chi$ around
$\phi^\star=-\sqrt{\frac{2}{3}}\ln \frac{t}{\tilde{c}}$ to get

\be V(\phi(t))=V(\phi^\star)(1+\alpha W_V(\phi^\star){V}(\phi) {t}^2)+h\label{expansionV}\ee and

\be \chi(\phi(t))=\chi(\phi^\star)(1+\alpha W_\chi(\phi^\star){V}(\phi) {t}^2)+h\label{expansionchi}\ee
where $\alpha$ is a constant. By substituting equations (\ref{expansionV}) and (\ref{expansionchi}) into the equations
(\ref{expansionH}, \ref{expansiondotphi}, \ref{expansionphi}, \ref{expansionrho}) the theorem is proven. $\square$

\subsection{A global singularity theorem}

Finally, we will state (without a rigorous proof) a global singularity theorem which is in some way an extension of
Theorem 6 in \cite{Foster1998sk} (page 3501). It is not totally an extension of this theorem, since in our framework it
is very difficult to prove that the correspondence with the massless minimally coupled scalar field cosmologies is
one-to-one.

The theorem states the following:

\begin{thm}\label{globalinitialsingularity}
Let be $V\in\mathcal{E}^2$ such that $N^2_{\pm\infty}<6$ and $\chi\in\mathcal{E}^2$ such that
\begin{enumerate}
\item[i)] $0<\gamma<\frac{4}{3}$ and $M_{\pm\infty}>-\frac{\sqrt{6}(\gamma-2)}{3\gamma-4}$ or
\item[ii)] $\frac{4}{3}<\gamma<2$ and $M_{\pm\infty}<-\frac{\sqrt{6}(\gamma-2)}{3\gamma-4}$
\end{enumerate}
Then, it is verified asymptotically that:

\begin{align}
&H= \frac{1}{3t} +O\left(\epsilon^\pm_V(t)\right),\\
&\phi= \pm\sqrt{\frac{2}{3}}\ln \frac{t}{\tilde{c}} + O\left(t\epsilon^\pm_V(t)\right),\\
&\dot\phi=\pm\sqrt{\frac{2}{3}} t^{-1}+ O\left(\epsilon^\pm_V(t)\right),
\\&\rho=\chi_0^2 t^{-\gamma}\chi\left(\pm\sqrt{\frac{2}{3}}\ln \frac{t}{\tilde{c}}\right)^{\frac{3\gamma}{2}-2}\left(1+ O\left(t\epsilon^\pm_V(t)\right)\right),\end{align} where $\epsilon^\pm_V(t)=t V\left(\pm\sqrt{\frac{2}{3}}\ln \frac{t}
{\tilde{c}}\right).$
\end{thm}

{\bf Sketch of the proof}

Following the reasoning \cite{Foster1998sk}, it is sufficient to
demonstrate that almost all the solutions are past asymptotic to
the critical point $P_1$ (in $\infty$ or in $-\infty$). Since $x$
is monotonic, it is sufficient to consider solutions in
$\Omega(x_0)\subset\Sigma(x_0)$ where $x_0$ is arbitrary. Since
$\Sigma(x_0)$ is compact and it contains its past orbits, then all
the points $p$ must have an $\alpha$-limit set, $\alpha(p)$.
Particularly, for the points in the physical space $\Omega(x_0),$
the theorem  \ref{Theorem IV} implies that  $\alpha(p)$ must
contain almost always a critical point with $\varphi=0$
($\phi=\pm\infty$). By the discussion in the section
\ref{estructura}, each point with $\varphi=0$ being a limit point
of the physical trajectory, must be part of the non-physical
boundary $\partial\Omega(x_0)$ and then must have $x=0.$ Since $x$
is monotonically increasing, the set $\alpha(p)$ must be contained
completely in the plane $x=0,$ or namely in $\partial\Omega(x_0).$
It can be proved that the only conceivable generic past attractor
are the critical points $P_1$ in $\pm\infty$ (the other critical
points cannot be generic sources by our previous linear analysis).

\section{Conclusions}

In this paper we have investigated models with additional
(non-gravitational) interaction between DE and DM. This kind of
interaction is justified if the interacting components are of
unknown nature, as it is the case for the DM and the DE, the
dominant components in the cosmic fluid. We have investigated
these models from the dynamical systems viewpoint. The functional
form of the potential and the coupling function is arbitrary from
the beginning. Some general results are obtained and proved by
considering general hypotheses on these input functions.

We have proved (by using normalized variables) that the scalar
field typically diverges into the past. This is formulated in
Theorem \ref{Theorem IV}. It is an extension of the theorem 1 in
the reference \cite{Foster1998sk} to the non-minimally coupled
scalar field setting.

In lemma \ref{Theorem I} it is proved that the orbit passing
through an arbitrary point $p\in\Sigma$ (representing cosmological solutions
 with non-vanishing dimensionless background
energy density and positive finite Hubble parameter) is past
asymptotic to a regime where the Hubble parameter
diverges containing an initial singularity into the past, and is
future asymptotic to a regime where the background
density is negligible into the future. This result is obtained
by constructing a monotonic function defined on an invariant set
and by applying the LaSalle monotonocity principle (theorem 4.12,
\cite{reza}).

We have proved the Theorem \ref{Theorem IV} that makes clear that
in order to investigate the generic past asymptotic behavior of
our system we must seek on the limit where the scalar field
diverges.

By assuming some regularity conditions on the potential and on the
coupling function in that regime we have constructed a dynamical
system (well suited to investigate the dynamics where the scalar
field diverges, i.e. near the initial singularity). The critical
points therein are investigated and the cosmological solutions
associated to them are characterized. We find the existence of
three critical points $P_3,$ $P_5$ and $P_6.$ They are in the
boundary of the phase space $\Sigma_\epsilon.$ They represent
cosmological scaling solutions (where the contribution of the
dimensionless potential energy is negligible). By tuning the free
parameters they can be accelerating. In contrast in the reference
\cite{Foster1998sk} there exists only one (in our notation, $P_4$)
representing an accelerating cosmology. The solutions associated
to $P_{1,2}$ ($p_\mp$ in the notation in \cite{Foster1998sk})
represent stiff and then decelerating solutions (actually
solutions associated to a massless scalar field).

We have proved a theorem (theorem \ref{initialsingularity}) which
is an extension of the theorem 4 in \cite{Foster1998sk} to the STT
framework. Also, we sketch the proof of the global singularity
theorem \ref{globalinitialsingularity}. Theorem
\ref{globalinitialsingularity} indicates that the past asymptotic
structure of non-minimally coupled scalar field theories with FRW
metric, as in the FRW general relativistic case, is independent of
the exact details of the potential and/or the details of the
background matter and the coupling function.

This is a conjecture with solid theoretical and numerical
foundations (see figures 1 and 2 in \ref{toy}). To prove
that the family of solutions which asymptotically approach $P_1$
are completely characterized by the solution space of the massless
scalar field cosmological model (i.e., $V$ and $\chi$ and then
$\rho$, being dynamically insignificant in the neighborhood of the
singularity $P_1$) it is required to prove that this
correspondence is one-to-one and continuous, which is hard to do
in our scenario.

\section{Acknowledgements}

G.L. wishes to thanks to I. Quiros for reading the original
manuscript and for helpful suggestions. A. A. Coley is
acknowledged for his comments and for provide us the reference
\cite{Billyard:2000bh}.  D. Gonzales and Y. Napoles are
acknowledged for invaluable assistance. Thanks to R. Cardenas for
its comments on the equivalence of the Jordan and Einstein frames
and for reviewing the English. Thanks to the two anonymous
referees for their criticism and comments. This investigation was
supported by MES of Cuba.

\appendix

\section{Some terminology and results from the Dynamical Systems Theory}\label{appendixA}

In the following $\phi(t,x)$ denotes the flow generated by the vector field (or differential equation)

\begin{equation}\dot x(t)=f(x(t)),\,x(t)\in\mathbb{R}^n.\label{vectorfield}\end{equation}

\subsection{Limit sets}

\begin{defn}[definition 8.1.1, \cite{wiggins} page 104]\label{omegalimitpoint} A point $x_0\in\mathbb{R}^n$ is called an $w$-limit point of
$x\in\mathbb{R}^n,$ denoted $\omega(x),$ if there exists a sequence $\{t_i\},\,t_i\rightarrow\infty$ such that
$\phi(t_i,x)\rightarrow x_0.$ $\alpha$-limits are defined similarly by taking a sequence $\{t_i\},\,t_i\rightarrow -\infty.
$

\end{defn}

\begin{defn}[definition 8.1.2, \cite{wiggins} page 105]\label{omegalimitset} The set of all $w$-limit points of a flow or map is called a
$w$-limit set. The $\alpha$-limit is similarly defined.

\end{defn}

\begin{prop}[proposition 8.1.3 \cite{wiggins}, page 105]\label{omegalimitsetproperties} Let $\phi_t(\cdot)$ be a flow generated by a vector field and
let $M$ be a positively invariant compact set for this flow (see
definition 3.0.3 page 28 \cite{wiggins}). Then for $p\in M,$ we
have

\begin{enumerate}
\item{i)} $\omega(p)\neq\emptyset$
\item[ii)] $\omega(p)$ is closed
\item[iii)] $\omega(p)$ is invariant under the flow, i.e., $\omega(p)$ is a union of orbits.
\item[iv)] $\omega(p)$ is connected.
\end{enumerate}

\end{prop}

A similar result follows for $\alpha$-limit sets provided the hypothesis of the proposition are satisfied for the time
reversed flow (see proposition 1.1.14 in \cite{wiggins90}).

\subsection{Center Manifolds}

A general vector field can be transformed locally in the
neighborhood of a fixed point into a vector field of the form

\bea\dot x=Ax+f(x,y),\\
\dot y=By+g(x,y),\, (x,y)\in\mathbb{R}^c\times\mathbb{R}^s,\label{vect}\eea where

\bea f(0,0)=g(0,0)=Df(0,0)=Dg(0,0)=0,\eea $A$ is a $c\times c$
matrix having eigenvalues with zero real parts, $B$ is a $s\times
s$ matrix with negative real parts, and $f$ and $g$ are $C^r$
functions ($r\geq 2$).

The Center Manifold for the vector field (\ref{vect}) is defined
as follows:

\begin{defn}[definition 18.1.1, \cite{wiggins} page 246]\label{centermanifold} An invariant manifold will be called a center manifold for
(\ref{vect}) if it can locally be represented as follows

$$W_{\text{loc}}^c(0)\left\{(x,y)\in \mathbb{R}^c\times\mathbb{R}^s| y=h(x), |x|<\delta, h(0)=0, Dh(0)=0\right\}$$ for
$\delta$ sufficiently small.

\end{defn}

The conditions $h(0)=0, Dh(0)=0$ imply that $W_{\text{loc}}^c(0)$
is tangent to $E^c$ at $(x,y)=(0,0),$ where $E^c$ is the center
subspace, i.e., the invariant set spanned by the eigenvectors
whose associated eigenvalues have zero real parts.

\begin{theorem}[Center Manifold (theorem 2.7.1 \cite{arrowsmith})]\label{centermanifoldtheorem} Let be $\phi_t$ the flow of a vector field, $\bf X,$
then, there exists locally a center manifold, $W_{\text{loc}}^c,$
containing the origin and invariant under $\phi_t$ such that
$W_{\text{loc}}^c,$ has tangent space $E^c$ at $x=0.$ This
manifold is $C^k$ for all $k\in\mathbb{N},$ but its domain of
definition can depend on $k.$ Furthermore, there are locally
smooth stable and unstable manifolds, $W_{\text{loc}}^s$ and
$W_{\text{loc}}^u,$ which contain $x=0,$ are invariant under
$\phi_t,$ have tangent spaces $E^s$ and $E^u,$ respectively, and
are such that $\phi_t|_{W_{\text{loc}}^s}$ is a contraction while
$\phi_t|_{W_{\text{loc}}^u}$ is an expansion.

\end{theorem}

\subsection{Monotone functions and Monotonicity Principle}

\begin{defn}[definition 4.8 \cite{reza}, page 93]\label{Definition 4.8} Let $\phi_t$ be a flow on $\mathbb{R}^n,$ let $S$ be an invariant set
of $\phi_t$ and let $Z:S\rightarrow$ be a continuous function. $Z$
is monotonic decreasing (increasing) function for the flow
$\phi_t$ means that for all $x\in S,$ $Z(\phi_t(x))$ is a
monotonic decreasing (increasing) function of $t.$

\end{defn}

\begin{prop}[Proposition 4.1, \cite{reza}, page 92]\label{Prop 4.1}

Consider a differential equation $x'=f(x)$, $x\in \mathbb{R}^n$ with
flow $\phi_t.$ Let $Z:\mathbb{R}^n\rightarrow \mathbb{R}$ be a $C^1\left(\mathbb{R}^n\right)$ function which satisfies
$Z'=\alpha Z,$ where $\alpha: \mathbb{R}^n\rightarrow \mathbb{R}$
is a continuous function. Then, the subsets of
$\mathbb{R}^n$ defined by  $Z>0,$ $Z=0,$ or $Z<0$ are invariant sets for $\phi_t.$

\end{prop}

\begin{theorem}[Monotonicity Principle, \cite{reza}, page 103]\label{theorem 4.12}

Let $\phi_t$ be a flow on $\mathbb{R}^n$ with $S$ an invariant
set. Let $Z: S\rightarrow\mathbb{R}$ be a
$C^1\left(\mathbb{R}^n\right)$ function whose range is the
interval $(a,\;b)$ where $a\in \mathbb{R} \cup \{-\infty\},$ $b\in
\mathbb{R} \cup \{+\infty\},$ and $a<b.$ If $Z$ is decreasing on
orbits in $S,$ then for all $x\in S$, $\omega(x)\subset\{s\in \bar
S-S|lim_{y\rightarrow s} Z(y)\neq b \}$ and
$\alpha(x)\subset\{s\in \bar S-S|lim_{y\rightarrow s} Z(y)\neq a
\}.$

\end{theorem}

\section{Regularity conditions of the potential and of the coupling function at infinity}\label{appendixB}

In this section we present rigorous statements of what we mean by
regularity conditions. We present some worked examples that we
shall use to construct a toy model. Our purpose is illustrated by
the techniques for analysis in the region $\phi=+\infty.$ In
principle the analysis is general enough to be applied to more
physically interesting situations.

\begin{defn}[see reference \cite{Foster1998sk}]\label{WBI}
Let $V:\mathbb{R}\rightarrow \mathbb{R}$ be a $C^2$ non-negative function. Let there exist
some $\phi_0>0$ for which $V(\phi)> 0$
 for all $\phi>\phi_0$ and some number $N$ such that the function
$W_V:[\phi_0,\infty)\rightarrow R$,
$$ W_V(\phi)=\frac{\partial_\phi V(\phi)}{V(\phi)} - N $$
 satisfies
\begin{equation}
\lim_{\phi\rightarrow\infty}W_V(\phi)=0.\label{Lim}
\end{equation}
Then we say that $V$ is Well Behaved at Infinity (WBI) of exponential order $N$.
\end{defn}

It is important to point out that $N$ may be 0, or even negative. Indeed the class of WBI functions of order 0 is of
particular interest, containing all non-negative polynomials as remarked in \cite{Foster1998sk}.

In that reference it was defined a procedure for classifying the smoothness of WBI functions at infinity. If we have some
coordinate transformation $\vphi=f(\phi)$ which maps a neighborhood of infinity to a neighborhood of the origin, then if
$g$ is a function of $\phi$,
 $\overline{g}$ is the function of $\vphi$ whose domain is the range of
$f$ plus the origin, which takes the values;

$$
\overline{g}(\vphi)=\left\{\begin{array}{rcr} g(f^{-1}(\vphi))&,&\vphi>0\\
                                     \lim_{\phi\rightarrow\infty} g(\phi)&,&\vphi=0 \end{array}\right.
$$

\begin{defn}[see reference \cite{Foster1998sk}]\label{CkWBI}
A $C^k$ function $V$ is class k WBI if it is  WBI and if there exists $\phi_0>0$ and a coordinate  transformation
$\vphi=f(\phi)$ which maps the interval $[\phi_0,\infty)$ onto
$(0, \epsilon]$, where $\epsilon=f(\phi_0)$ and $\lim_{\phi\rightarrow\infty} f=0$, with  the following additional
properties:
\begin{tabbing}
i)\hspace{0.4cm}\=  $f$ is $C^{k+1}$ and strictly decreasing.\\
ii)            \>the functions $\overline{W}_V(\vphi)$ and $\overline{f'}(\vphi)$ are $C^k$ on
the \\ \> closed interval $[0,\epsilon]$.\\
iii)           \> ${\displaystyle \frac{d\overline{W}_V}{d\vphi}(0)=\frac{d\overline{f'}}{d\vphi}(0)=0.}$
\end{tabbing}
\end{defn}
We designate the set of all class k WBI functions ${\cal E}^k_+.$ In the table 1 of \cite{Foster1998sk} is incorporated
a diverse range of qualitative behavior within the framework of ${\cal E}^k_+$ potentials.

\subsection{Worked examples.}

\subsubsection{Power-law coupling function.}

Let us consider the coupling function \be \chi(\phi)=\left(\frac{3\alpha}{8}\right)^{\frac{1}{\alpha}}\chi_0(\phi-\phi_0)^
{\frac{2}{\alpha}},\; \alpha>0,\text{const.},\,\phi_0>0.\label{couplingexample}\ee

Observe that $$\frac{d\ln \chi(\phi)}{d\chi}=\frac{2}{\alpha(\phi-\phi_0)}\neq 0$$ for all finite value of $\phi.$ Then,
the critical point $Q$ located at the invariant set $X^0$ does not exist for this choice of coupling function. By our
analysis in section \ref{Qualitative}, the early time dynamics is associated to the limit where the scalar field diverges.

This choice produces a coupling BD parameter given by

$$2\omega(\chi)+3=\frac{4}{3}\alpha\left(\frac{\chi}{\chi_0}\right)^\alpha.$$

This types of power law couplings were investigated in
\cite{DTorres} from the astrophysical viewpoint. For scalar-tensor
theories without potential, the cosmological solutions for the
matter domination era (in a Robertson-Walker metric) are
$a(t)\propto (\ln t)^{(\alpha-1)/3\alpha} t^{\frac{2}{3}},\;
\phi(t)\propto (\ln t)^{\frac{1}{\alpha}}.$ The values of the
parameter $\alpha$ in concordance with the predictions of ${}^4 H$
are $\alpha=1, \, 0.33,\, 3$ (see table 4.2 in \cite{DTorres}).

\subsubsection{The Albrecht-Skordis potential.}

Albrecht and Skordis \cite{Albrecht1999rm} have proposed a particularly attractive model of quintessence. It is driven by
a potential which introduces a small minimum to the exponential potential:
\begin{equation}
V(\phi )=e^{-\mu \phi }{\left( A+(\phi -B)^2\right).}  \label{Albrecht-Skordis}
\end{equation}
Unlike previous quintessence models, late-time acceleration is
achieved without fine tuning of the initial conditions. The
authors argue that such potentials arise naturally in the
low-energy limit of $M$-theory. The constant parameters, $A$
 and $B$, in the potential take values of order $1$ in Planck units, so there is also no fine tuning of the potential
 (we suppose also that $\mu\neq 0$). They show that, regardless of the initial conditions, $\rho _\phi $ scales, with
 $\rho \propto \rho _\phi \propto t^{-2}$ during the radiation and matter eras, but leads to permanent vacuum domination
 and accelerated
expansion after a time which can be close to the present.

The extremes of of the potential (\ref{Albrecht-Skordis}) are located at $\phi^{\pm}=\frac{1+B\mu-\sqrt{1-A\mu^2}}{\mu}.$
They are real if $1\geq
\mu ^2A.$ The local minimum (respectively, local maximum)  is located at $\phi^-$ (respectively $\phi^+$) since
$$\pm V''(\phi^{\pm})=- 2 V_0\sqrt{1-A\mu^2}e^{-\left(1+B\mu\pm\sqrt{1-A\mu^2}\right)}<0.$$

As we investigated in section \ref{Qualitative}, the late time
dynamics in the invariant set $Z^0$ is associated with the
extremes of the potential. When we restrict ourselves to this
invariant set, we find that the critical point associated to
$\phi^+$ is always a saddle point of the corresponding phase
portrait. The critical point associated to $\phi^-$ could be
either a stable node or a stable spiral if
$$\frac{8(3+2\mu^2)}{\left(3+4\mu^2\right)^2}<A\leq
\frac{1}{\mu^2}$$ or
$$A<\frac{8(3+2\mu^2)}{\left(3+4\mu^2\right)^2}.$$ The early time
dynamics in this invariant set corresponds to the limit
$\phi=+\infty.$

Let us concentrate now in how to apply the mathematics introduced
in this section.

Observe first that \be W_\chi(\phi)=\partial_\phi \chi(\phi)/\chi(\phi)=\frac{2}{\alpha  (\phi -\phi_0)} \Rightarrow
\lim_{\phi\rightarrow +\infty} W_\chi(\phi)=0\ee and  \be W_V(\phi)=\partial_\phi V(\phi)/V(\phi)+\mu=\frac{2(\phi-B)}{A+(B-\phi)^2} \Rightarrow \lim_{\phi\rightarrow +\infty} W_V(\phi)=0.\ee  In other words, the coupling function (\ref{couplingexample}) and the potential (\ref{Albrecht-Skordis}) are WBI of exponential orders $M=0$ and $N=-\mu,$ respectively.

It is easy to prove that Power-law coupling and the Albrecht-Skordis potential are at least  ${\cal E}^2_+,$  under the
admissible coordinate transformation \be \vphi=\phi^{-1}=f(\phi)\label{transformAS}.\ee

Using the above coordinate transformation we find

\be\overline{W}_{\chi}(\vphi)=\left\{\begin{array}{rcr} \frac{2 \varphi }{\alpha  (1-\varphi  \phi_0)}&,&\vphi>0\\
                                     0&,&\vphi=0 \end{array}\right.\label{WchiAS}\ee

\be\overline{W}_V(\vphi)=\left\{\begin{array}{rcr} -\frac{2\vphi(B\vphi-1)}{A\vphi^2+(B\vphi-1)^2}&,&\vphi>0\\
                                     0 &,&\vphi=0 \end{array}\right.\label{WVAS}\ee
and

\be\overline{f'}(\vphi)=\left\{\begin{array}{rcr} -\vphi^2&,&\vphi>0\\
                                     0&,&\vphi=0 \end{array}\right.\label{fAS}\ee

\subsection{The dynamics in the limit $\phi\rightarrow +\infty.$ An example.}\label{toy}

We will present here a toy model by choosing a Power-law coupling
and Albrecht-Skordis potential with the only purpose of illustrate
our previous results. Due to the generality of our study it can be
applied for more realistic scalar-tensor cosmologies.

Let us consider a toy model with the coupling function (\ref{couplingexample}) and the potential (\ref{Albrecht-Skordis}).
 Observe that this toy model corresponds to a theory written in the Jordan frame with action
\begin{align}& S_{JF}=\int_{M_4} d{ }^4 x \sqrt{|\bar{g}|}\left\{\frac{1}{2}\chi \bar{R}-\frac{1}{2}\frac{\omega(\chi)}
{\chi}(\bar{\nabla}\chi)^2-\bar{V}(\chi)
+\mathcal{L}_{matter}(\mu,\nabla\mu,\bar{g}_{\alpha \beta})\right\}
\end{align}
with BD coupling $$\omega(\chi)=-\frac{3}{2}+\frac{2}{3}\alpha\left(\frac{\chi}{\chi_0}\right)^\alpha,$$ and potential $$\bar{V}(\chi)=\chi^2 e^{-\mu \phi(\chi) }{\left( A+\left(\phi(\chi) -B\right)^2\right)}$$ where $$\phi(\chi)=\phi_0+\sqrt{\frac{8}{3\alpha}}\left(\frac{\chi}{\chi_0}\right)^{\frac{\alpha}{2}}.$$ We have considered also, conformal metric $$\bar{g}_{\alpha \beta}=\chi(\phi)^{-1}g_{\alpha \beta}.$$

\begin{figure}\label{FIG1}
\begin{center}
\hspace{0.4cm}
\put(120,0){${P_4}$}
\put(133,0){${P_2}$}
\put(0,0){${P_1}$}
\includegraphics[width=5cm, height=4cm]{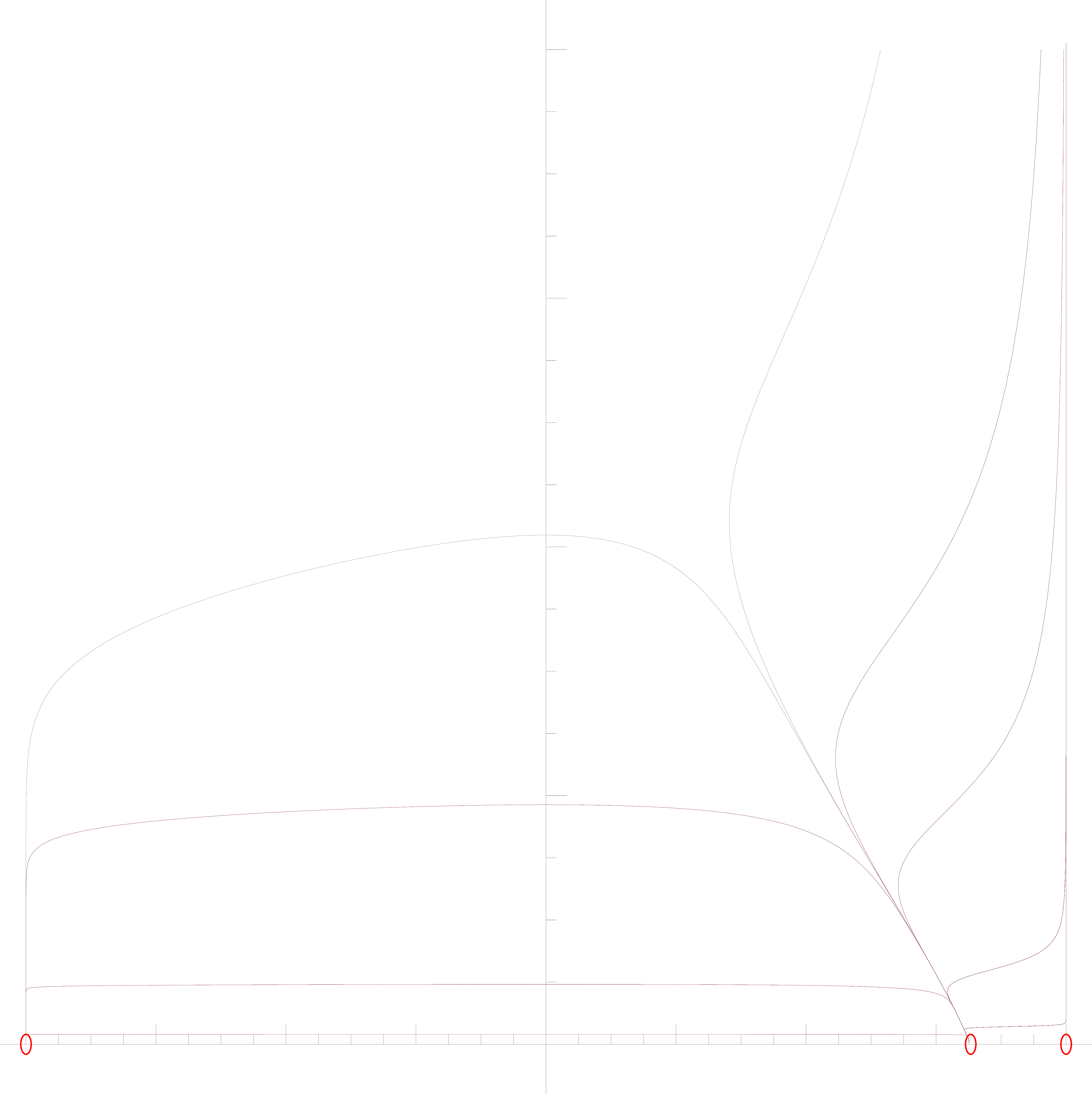}
\caption{Orbits in the invariant set $\{z=0\}\subset \bar{\Sigma}_\epsilon$ for the model with coupling function (\ref{couplingexample})  potential (\ref{Albrecht-Skordis}). We select the values of the parameters: $\epsilon=1.00,$ $\mu= 2.00, A = 0.50, \alpha = 0.33, B = 0.5,$ and $\phi_0=0.$ Observe that i) almost all the orbits are past asymptotic to $P_1;$ ii) $P_2$ is a saddle, and iii) the center manifold of $P_4$ attracts all the orbits in the $\{z=0\}$. However, it is no more an attractor in the invariant set $z>0,\,\varphi=0$ (see figure 2 in section \ref{toy}).}
\end{center}
\end{figure}

\begin{figure}
\begin{center}
\hspace{0.4cm}
\put(110,57){${P_5}$}
\put(120,0){${P_4}$}
\put(65,108){${P_3}$}
\put(133,0){${P_2}$}
\put(0,0){${P_1}$}
\includegraphics[width=5cm, height=4cm]{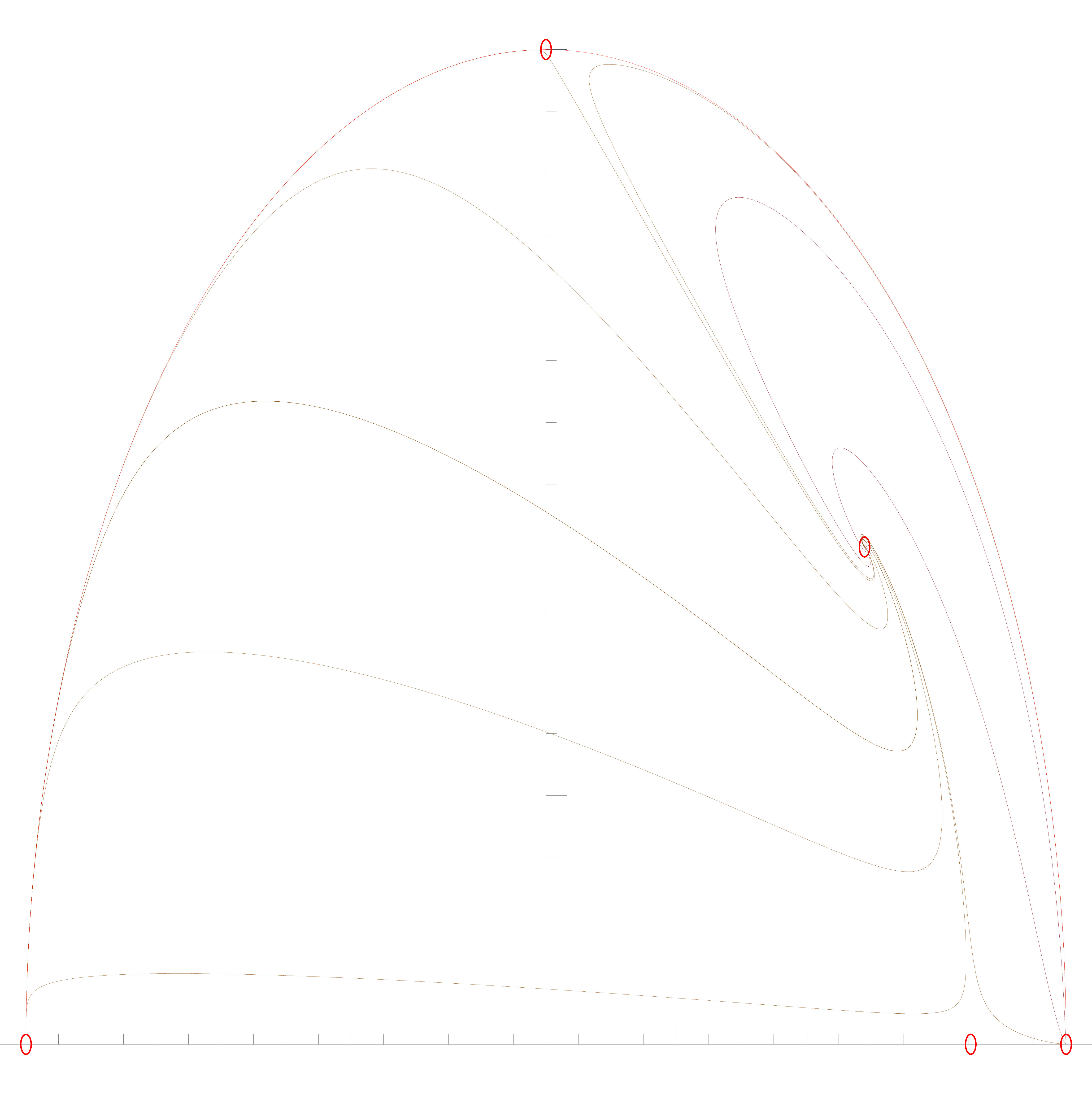}
\label{FIG2}
\caption{Orbits in the invariant set $\{\varphi=0\}\subset \bar{\Sigma}_\epsilon$ for the model with coupling function
(\ref{couplingexample})  potential (\ref{Albrecht-Skordis}). We select the values of the parameters: $\epsilon=1.00,$
$\mu= 2.00, A = 0.50, \alpha = 0.33, B = 0.5,$ and $\phi_0=0.$ In the figure i) $P_{1,2}$ are local past attractors,
but $P_1$ is the global past attractor; ii) $P_{3,4}$ are saddles, and iii) $P_5$ is a local future attractor.}
\end{center}
\end{figure}
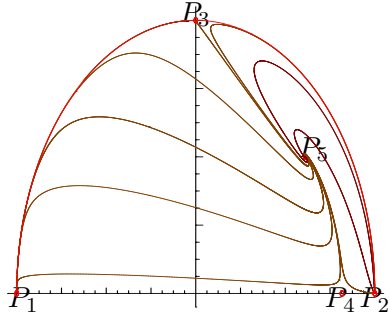

In this example, the evolution equations for $y,$ $z,$ and $\vphi$ are given by the equations
(\ref{eqinfy}-\ref{eqinfphi}) with $M=0,$ $N=-\mu$ and $\overline{W}_{\chi},$ $\overline{W}_V,$ and $\overline{f'},$
given respectively by (\ref{WchiAS}), (\ref{WVAS}) and (\ref{fAS}). The state space is defined by
$\overline{\Sigma}_\epsilon=\left\{(y,\,z,\,\vphi): 0\leq y^2+z^2\leq 1,\,0\leq\varphi\leq \epsilon\right\}.$

The critical points of the system (\ref{eqinfy}-\ref{eqinfphi}) in this example are $P_{1,2}=(\mp 1,0,0)$, $P_3=\left
(0,1,0\right),$ and $P_4=\left(\frac{\mu}{\sqrt{6}},0,0\right)$, and $P_{5,6}=\left(\sqrt{\frac{3}{2}}\frac{\gamma}{\mu}
,\mp\frac{\sqrt{-12\gamma+4\mu^2}}{2\mu},0\right).$ The points $P_{1,2,3}$ exist for all the values of the free parameters.
 The critical point $P_4$ exists for $\mu^2\leq 6.$ The critical point $P_5$ exists if $\mu\leq -\sqrt{3\gamma}$ whereas
 the critical point $P_6$ exists if $\mu\geq \sqrt{3\gamma}.$ We will characterize the critical points $P_{5,6}$ in more
 detail (for the analysis of the other critical points we submit the reader to table \ref{crit}). The critical points
 $P_{5,6}$ corresponds to those studied in the book \cite{Coley:2003mj} (see equation 4.23 p 49) with the identifications
 $\Psi=y$ and $\Phi^2=\frac{V(\phi)}{3 H^2}={\frac{3\gamma(2-\gamma)}{2\mu^2}}$ and $k=-\mu.$ As stated in that
 reference the scalar field 'inherits' the equation of state of the fluid, i.e., $\gamma_\phi=\gamma$. Then this solutions
 represent cosmological kinetic-matter scaling solutions \footnote{See reference \cite{Holden:1999hm} for a notion of
 'scaling' solutions, particularly, kinetic-matter scaling solutions.} (the potencial energy density is negligible).
 These critical points represent accelerating cosmologies for $0<\gamma<\frac{2}{3}.$ The eigenvalues of the matrix of
 derivatives evaluated at $P_{5,6}$ are $\left(0,-\frac{2-\gamma}{4\mu}\pm \frac{1}{4\mu}\sqrt{(2-\gamma)\left(24\gamma^2+
 \mu^2(2-9\gamma)\right)}\right).$ The orbits initially in the stable subspace of $P_{5,6}$ spiral-in around $P_{5,6}$ if
 $\mu^2>24\gamma^2/(-2+9\gamma)$ provided $\frac{2}{9}<\gamma<2,\gamma\neq \frac{4}{3}.$ Otherwise $P_{5,6},$ looks like
 a stable node for the orbits lying in the stable subspace. The center subspace is tangent to the critical points in the
 direction of the $\vphi$ axis.


\section*{References}


\begin{thebibliography}{99}

\bibitem{sncmbsdss}
  Riess A.~G.~{\it et al.}, 2004,  
  Astrophys.\ J.\  {\bf 607}, 665;
Astier P. {\it et al.}, 2006,
  Astron.\ Astrophys.\  {\bf 447}, 31.
Riess A.~G. {\it et al.}, 2006,
  arXiv: astro-ph/0611572;
  Spergel D. N. {\it et al.}, 2003,  
  Astrophys.\ J.\ Suppl.\  {\bf 148}, 175;
  Goldstein J.~H. {\it et al.}, 2003,
  %
  Astrophys.\ J.\  {\bf 599}, 773; 
Readhead A.~C.~S.{\it et al.}, 2004,
  Astrophys.\ J.\  {\bf 609}, 498;
  Tegmark M. {\it et al.}, 2004,  
  %
  Phys.\ Rev.\ D {\bf 69}, 103501;
Percival W.~J. {\it et al.}, 2006,
  arXiv: astro-ph/0608635.

\bibitem{Brans:1961sx}
  Brans C. and Dicke R.~H., 1961,
  Phys.\ Rev.\  {\bf 124}, 925.

\bibitem{STT} Bekeinstein J. D., 1977, Phys. Rev. D {\bf 15}, 1458; Bergman P. G., 1968 Int. J. Theor. Phys. {\bf 1}, 25; Nordtvedt K., 1970, Astrophys. J. {\bf 161}, 1059; Wanoger R. V., 1970, Phys. Rev. D {\bf 1} 3209.

\bibitem{Green:1996bh}
  Green M.~B., Hull C.~M. and Townsend P.~K., 1996,
  Phys.\ Lett.\  B {\bf 382}, 65.

\bibitem{Gonzalez:2004dh}
 Gonzalez A., Matos T. and Quiros I., 2005,
  Phys.\ Rev.\  D {\bf 71}, 084029.


\bibitem{quintessence} Caldwell R.~R., Dave R. and Steinhardt P.~J., 1998,
Phys.\ Rev.\ Lett.\ {\bf 80}, 1582;
Kolda C. F. and Lyth D. H., 1999, \PL \textbf{B458}, 197;
Sahni V., 2002, \CQG \textbf{19}, 3435;
Padmanabhan T., 2003, \PRTS \textbf{380}, 235.

\bibitem{SahniCopeland:2006wr} V.~Sahni and A.~Starobinsky, 2006,
  Int.\ J.\ Mod.\ Phys.\  D {\bf 15}, 2105
E.~J.~Copeland, M.~Sami and S.~Tsujikawa, 2006,
  Int.\ J.\ Mod.\ Phys.\  D {\bf 15}, 1753.
\bibitem{BarrowandParsons} Barrow J. D. and Parsons P., 1997, Phys. Rev. D {\bf 55}, 1906; Will C. M., 1993, Theory and Experiment in Gravitational Physics, Cambridge University Press, Cambridge.

\bibitem{doran} Doran M. and Jaeckel J., 2002, Phys. Rev. D {\bf 66} 043519.


\bibitem{solarsystem} Abramovici A.  et al, 1992, Science {\bf 256}, 325; Bradaschia C. et al., 1990, Nucl. Instrum. and Methods A {\bf 289}, 518; S. Buchmann et al., 1996, in Proccedings of the Seventh Marcel Grossman Meeting on General Relativity, ed. Jantzen R. T. and Keiser G. M., World Scientific, Singapore; Hough J. et al, 1993, in  Proccedings of the Sixth Marcel Grossman Meeting on General Relativity, Singapore.

\bibitem{Faraoni:1998qx}
 Faraoni  V., Gunzig  E. and Nardone P., 1999,
  Fund.\ Cosmic Phys.\  {\bf 20}, 121.
  [arXiv:gr-qc/9811047].

\bibitem{nucleosinthesis} Barrow J. D., 1997, Phys. Rev. D {\bf 35}, 1805; Serna A. and Alimi J. M., 1996, Phys. Rev. D {\bf 53}, 3087.


\bibitem{Kaloper1997sh}
Kaloper N. and Olive K.~A., 1998,
  Phys.\ Rev.\  D {\bf 57}, 811.

\bibitem{Coley:2003mj}
  Coley A.~A., 2003,
  Dynamical systems and cosmology,
{\it  Dordrecht, Netherlands: Kluwer.}

\bibitem{equivalent} Catena R., Pietroni M. and Scarabello L., 2007,
Phys.\ Rev.\ D {\bf 76}, 084039.
Faraoni V. and Lanahan-Tremblay N.,
arXiv:0808.1943 [gr-qc].

\bibitem{quantum} Flanagan  E.E., 2004, Class. Quantum Grav. {\bf 21}, 3817; Faraoni V. and Nadeau S., 2007, Phys. Rev. D {\bf 75},
023501.

\bibitem{Gonzalez2006cj}
 Gonzalez T., Leon G. and Quiros I., 2006,
  Class.\ Quant.\ Grav.\  {\bf 23}, 3165.
  
\bibitem{trodden}
  Trodden M. and Carroll S.~M., 2004,
  arXiv:astro-ph/0401547.

\bibitem{Billyard:2000bh}
  Billyard A. P. and Coley, A. A., 2000,
  Phys.\ Rev.\  D {\bf 61}, 083503
  [arXiv:astro-ph/9908224].

\bibitem{Boehmer:2008av}
  Boehmer C. G., Caldera-Cabral G., Lazkoz R. and Maartens R., 2008
  Phys.\ Rev.\  D {\bf 78}, 023505

\bibitem{viableSTT} Tsujikawa S., Uddin K., Mizuno S., Tavakol R. and Yokoyama J., 2008,
Phys.\ Rev.\ D {\bf 77}, 103009.



\bibitem{coley2} Coley A. A., 1999, Gen. Rel. Grav. {\bf 31}, 1295; Mimoso J. P. and Wands D., 1995, Phys. Rev. D {\bf 51}, 477.

\bibitem{dynamical} Tocchini-Valentini D. and Amendola L., 2002,
Phys.\ Rev.\ D {\bf 65}, 063508;
Amendola L., 2000
Phys.\ Rev.\ D {\bf 62}, 043511.



\bibitem{Foster1998sk}
  Foster S., 1998,
  Class.\ Quant.\ Grav.\  {\bf 15}, 3485.

\bibitem{Miritzis2003ym}
  Miritzis J., 2003,
  Class.\ Quant.\ Grav.\  {\bf 20}, 2981.


\bibitem{arbitrary} Miritzis J., 2005, J. Math. Phys. {\bf 46}, 082502; Rendall A.D., 2004, Class. Quant. Grav. {\bf 21}, 2445; Rendall A.D., 2007, Class. Quant. Grav. {\bf 24}, 667.

\bibitem{Hertog} Hertog T., 2006, Phys. Rev. D {\bf 74}, 084008.

\bibitem{quiros}  Gonzalez T. and Quiros I., 2008, Class.\ Quant.\ Grav.\  {\bf 25}, 175019;
Curbelo R., Gonzalez T., Leon G. and Quiros I., 2006, Class.\ Quant.\ Grav.\  {\bf 23}, 1585.


\bibitem{Chimento2003iea}
 Chimento  L.~P., Jakubi A.~S., Pavon D.~ and Zimdahl  W., 2003,
  Phys.\ Rev.\  D {\bf 67}, 083513.

\bibitem{Wainwright2004cd}
  Wainwright J. and Lim W.~C., 2005,
  J.\ Hyperbol.\ Diff.\ Equat.\  {\bf 2}, 437.

\bibitem{reza} Tavakol  R., 1997, in {Dynamical Systems in Cosmology}, edited by Wainwright J.  and Ellis G.F.R. (Cambridge University Press, Cambridge).

\bibitem{Daniaetal} Gonzalez D. and Napoles Y., 2008, B. Sc. thesis, UCLV.

\bibitem{DTorres} Torres, D. F., PhD thesis, 1998.

\bibitem{Albrecht1999rm}
 Albrecht  A.~ and Skordis C., 2000,
  Phys.\ Rev.\ Lett.\  {\bf 84}, 2076.


\bibitem{Holden:1999hm}
  Holden D.~J.~ and Wands D., 2000,
  Phys.\ Rev.\  D {\bf 61}, 043506;
  Copeland E.~J., Liddle A.~R. and Wands D., 1998
Phys.\ Rev.\ D {\bf 57}, 4686.

\bibitem{wiggins} Wiggins S., 2003, ``Introduction to Applied Nonlinear Dynamical Systems and Chaos'', Springer.

\bibitem{wiggins90} Wiggins S., 1990, ``Introduction to Applied Nonlinear Dynamical Systems and Chaos'', Springer-Verlag, New York.

\bibitem{arrowsmith} Arrowsmith D. K. and Place C. M., 1990, ``An introduction to Dynamical Systems'', Cambridge University Press.

\end{thebibliography}
\end{document}